\documentclass[a4paper,aps,pra,twocolumn,superscriptaddress,showpacs]{revtex4}
\usepackage{graphicx}
\usepackage{amsmath}
\usepackage{amsfonts}
\usepackage{color}
\usepackage{bm}% bold math
\usepackage{pdfcomment}
\usepackage{pgfplots}
\usepackage{subfigure}

\newcommand{\media}[1]{\langle #1 \rangle}

\newcommand{\br}{\boldsymbol{r}}
\newcommand{\bp}{\boldsymbol{p}}

\newcommand{\comment}[1]{}

\usepackage{pgfplots}
\usetikzlibrary{plotmarks}

\begin{document}

%\preprint{}

\title{Quantum reservoirs with ion chains}

%--------------------------------------

\author{B. G. Taketani$^*$}
\affiliation{Theoretische Physik, Universit\"at des Saarlandes, D-66123 
Saarbr\"ucken, Germany}

\author{T. Fogarty\footnotemark}
\affiliation{Theoretische Physik, Universit\"at des Saarlandes, D-66123 
Saarbr\"ucken, Germany}
\affiliation{Physics Department, University College Cork, Cork, Ireland}
\affiliation{Quantum Systems Unit, OIST Graduate University, Okinawa, Japan}

%--------------------------------------
\author{E. Kajari}
\affiliation{Theoretische Physik, Universit\"at des Saarlandes, D-66123 
Saarbr\"ucken, Germany}

%---------------------------------------
\author{Th. Busch}
\affiliation{Quantum Systems Unit, OIST Graduate University, Okinawa, Japan}
%--------------------------------------
\author{Giovanna Morigi}
\affiliation{Theoretische Physik, Universit\"at des Saarlandes, D-66123 
Saarbr\"ucken, Germany}
%\affiliation{Departament de Fis\'ica, Universitat Aut\`onoma de Barcelona, 
%E-08193 Bellaterra, Spain}

\date{\today}

\begin{abstract}
Ion chains are promising platforms for studying and simulating quantum reservoirs. One interesting feature is that their vibrational modes can mediate entanglement between two objects which are coupled through the vibrational modes of the chain. In this work we analyse entanglement between the transverse vibrations of two heavy impurity defects  embedded in an ion chain, which is generated by the coupling  with the chain vibrations. We verify general scaling properties of the defects dynamics and demonstrate that entanglement between the defects can be a stationary feature of these dynamics. We then analyse entanglement in chains composed of tens of ions and propose a measurement scheme which allows one to verify the existence of the predicted entangled state.
\end{abstract}

%\keywords{}

\maketitle
\par
\footnotetext{Both authors contributed equally to this work. }
%%%%%%%%%%%%%%%%%%%%%%%%%%%%%%%%%%%%%%%%%%%%%%%
%%%%%%%%%%%%%%%%%%%%%%%%%%%%%%%%%%%%%%%%%%%%%%%
\section{Introduction}
Irreversibility characterises everyday life and is closely connected to the disappearance of quantum mechanical features in macroscopic objects. Microscopic descriptions of noise and dissipation usually start from a fully quantum mechanical model and have been discussed over the years in several seminal works \cite{Einstein,Chandrasekhar,FordKacMazur,Rubin:1963,Ullersma,Weiss:1999,Zurek:2003}. Recently, several studies have focused on thermalization in closed quantum systems \cite{Kinoshita:2004,Popescu:2006,Goldstein:2006,Rigol:2008,Riera:2012,Schmiedmayer,Polkovnikov} paying particular attention to the dynamics of many-body systems where thermalization does not occur \cite{Kinoshita:2004,Schmiedmayer}.

%%%%%%%%%%%%%%%%%%%
\begin{figure}[t]
\begin{center}
\includegraphics[angle=0,width=\columnwidth]{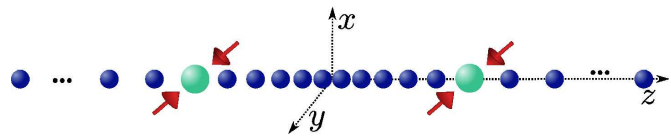}
\caption{(Color online) Two heavy ions are embedded in a linear chain of lighter ions.  The transverse modes of the heavy ions are coupled to the axial modes of the chain by means of an external force (such as a dipole force). This  can lead to the creation of robust entanglement between the transverse modes in the stationary state of the system.}
\label{fig:model}
\end{center}
\end{figure}
%%%%%%%%%%%%%%%%%%%

In this manuscript we consider a specific closed many-body system, which consists of a chain of ions in a linear Paul trap that couple via the repulsive Coulomb interaction \cite{Dubin:1999,Birkl:1992}. Two impurity defects, here two ions of larger mass, are placed within the chain as illustrated in Fig.~\ref{fig:model}, and couple with one another via the axial vibrations of the ion chain. In our model the displacement of an ion from its equilibrium position is described by quadratic terms in the expansion of the Coulomb potential. In this regime the dynamics are integrable and recurrences are observed on time scales determined by the finite system size. However,  since the coupling between the axial and transversal vibrations is a tunable parameter, the time scales between the appearance of (quasi) thermalization and recurrence are well separated and as a result the chain effectively acts as a thermal reservoir for each component. This behaviour agrees with the predictions of previous works, which studied thermalization of a component of a harmonic chain with nearest-neighbour interactions \cite{Rubin:1963,FordKacMazur,Ullersma,Kajari:2012}.

In the precursors of this work we have shown that the vibrational modes of a chain can also mediate entanglement between two impurity defects embedded in the chain. This entanglement is protected by spatial symmetries of the chain, it is robust  against variations in the chain size and can survive for times of the order of the recurrence time  \cite{Wolf:2011,Kajari:2012}. In Ref. \cite{Fogarty:2013} it was also found in a chain of ions interacting with the Couloumb long-range repulsion, assuming that the ions were uniformly spaced. Even if it exhibits several analogies, the entanglement dynamics found here cannot be put in direct connection with the ones in a chain with nearest-neighbour coupling. Moreover, typically the ion chain is realised in linear Paul traps, which impose a non-uniform density distribution: Bloch theorem does not apply. Finally, the dynamics discussed in Refs.~\cite{Wolf:2011,Kajari:2012,Fogarty:2013} are strictly valid for chains of hundreds of ions: only in this limit is the recurrence time sufficiently long so that entanglement is a (quasi) stationary feature. In experiments, however, the number of particles is typically of several tens to a maximum of a hundred, it is thus legitimate to ask whether such dynamics could be observed. The purpose of this work is to address these open questions: it elucidates what is the underlying mechanism which generates entanglement between two heavy impurity defects in an ion chain, it presents a systematic study of the parameter regimes under which entanglement is found, and identifies its feature in chains of tens of ions, which are at the borderline of the validity of the theory discussed in \cite{Wolf:2011,Kajari:2012,Fogarty:2013}. 

This work is organized as follows. In Sec.~\ref{Sec:MicroModel} we briefly review the basic features of entanglement generation in models of coupled oscillators with nearest-neighbour interactions. We then move to consider the experimentally realisable system of a linear chain of ions containing two defects in Sec.~\ref{sec:model} and characterise such a chain as a quantum reservoir.  The  entanglement between the two defect ions is numerically investigated in Sec.~\ref{sec:Results} as a function of the chain size, of the initial squeezing, and of the distance between the defects. We analyze the dynamics for small chains numerically and propose a measurement scheme for the defect states. Finally, the conclusions are drawn in Sec.~\ref{sec:conclusions}.

\section{Stationary entanglement in a chain of oscillators\label{Sec:MicroModel}}

In this section we briefly review the features of a simple microscopic model that allows one to study the entanglement generation between two oscillators via the interaction with a reservoir.  The theory presented here has been extensively discussed in Refs. \cite{Wolf:2011,Kajari:2012}. Elaborating from the knowledge developed in these previous works, we show with simple equations how normal modes, whose oscillations are localized at the defects positions, play an important role in entangling them.  These concepts will be important in order to understand the dynamics observed for ions interacting with long-range Coulomb repulsion. 

\subsection{Ion chain with nearest-neighbour coupling}
\label{micromodel}
We consider a chain of $N+2$ oscillators that couple with nearest-neighbour interaction. Among these, $N$ oscillators have mass $m$ and form a homogeneous linear chain with interparticle distance $a$ and coupling strength $\kappa$. The two additional defects have mass $M$ and are confined by a harmonic potential with trap frequency~$\Omega$. We denote $X_\mu$ as the position of the defect particles ($\mu=1,2$), and $x_i$ as the displacement of the chain particles from their respective equilibrium positions $x_i^ {(0)}=ia$. Here $i=\pm 1,\ldots,\pm A$ with $A=N/2$ for $N$ even, and $i=0, \pm 1,\ldots,\pm A$ with $A=(N-1)/2$ for $N$ odd. The corresponding canonically-conjugated momenta are $P_\mu$ and $p_i$, with non-vanishing commutation relations $[X_\mu,P_\mu]={\rm i}\hbar$ and $[x_i,p_i]={\rm i}\hbar$. The defects couple with the same strength $\gamma$ to the oscillators at positions $x_n$ and $x_{-n}$.

The Hamiltonian determining the dynamics of the closed system can then be written as
\begin{equation}
\label{eq:H}
{H = H_S+H_B+H_I}\,,
\end{equation}
where the free Hamiltonians for the two defect oscillators ({\it the system}) and for the $N$ chain oscillators ({\it the reservoir}) are given by 
\begin{align}
H_S&=\sum_{\mu=1}^2 \left[ \frac{P^2_\mu}{2 M} +\frac 12 M\Omega^2 X^2_\mu \right],
\label{eq:HSF}\\
H_B&=\sum_{i=1}^{N}\left[\frac{p_i^2}{2 m}+\frac{m}{2}\omega^2\, x_i^2\right]
+\frac{\kappa}{2}\sum_{i=1}^{N-1}(x_i- x_{i+1})^2\,.
\label{eq:HBF}
\end{align}
Here $\omega$ denotes the frequency of a harmonic potential of the Paul trap. The interaction Hamiltonian, which couples the system and bath oscillators is assumed to be instantly switched on at $t=0$ and for $t>0$ takes the form 
\begin{equation}
H_I=\frac\gamma 2 \Big[(X_1-x_{n})^2+(X_2-x_{-n})^2\Big]\,.
\label{eq:HIF}
\end{equation}
In the following we will assume $N$ to be odd, which is a convenient choice for chains of finite length and which does not effect the dynamics occurring in the chain bulk. We note that the latter can therefore also be used to characterize the thermodynamic limit $N\to\infty$.

In the presence of only one defect oscillator the above model is a generalization of models previously discussed by Rubin~\cite{Rubin:1963} and Ullerma~\cite{Ullersma}. They showed that a chain, which is initially prepared in a thermal state, can act as a thermal bath for a single defect under conditions which involve the mass ratio $M/m$, the strength of the coupling, and the time scales over which the dynamics are analyzed.  In the following we will discuss how this effect is modified in the presence of two defects. For this purpose it is convenient to recast the Hamiltonian in a coordinate system which highlights the symmetry properties of the dynamics. 

\subsection{Localized modes and entanglement}
\label{sec:loc_modes}
Let us introduce centre of mass (COM) and relative coordinates for the defect and chain particles as
\begin{align}
X_\pm &=(X_1\pm X_2)/\sqrt{2}\,,\\
x_{j,\pm}&=(x_j\pm x_{-j})/\sqrt{2}\,,
\end{align}
where $\pm$ indicates even or odd parity under mirror reflection about the chain centre, $x_0^{(0)}$. The corresponding canonically conjugate momenta are $P_\pm=(P_1\pm P_2)/\sqrt{2}$ for the defects and $p_{j,\pm}=(p_j\pm p_{-j})/\sqrt{2}$ for the chain oscillators. With this representation, and assuming that the defects are within the bulk of the chain and finite-size effects can be neglected, the Hamiltonian~\eqref{eq:H} can be written as $H= H_++H_-$, where $$H_\pm=H_{S,\pm}+H_{B,\pm}+\gamma X_\pm x_{j,\pm}\,,$$
and 
\begin{eqnarray}
H_{B,\pm}&=&\sum_{j_\pm=0}^A\left[\frac{p_{j,\pm}^2}{2 m}+\frac{m}{2}\omega^2\, x_{j,\pm}^2\right] \\
&&+ \sum_{j_\pm=0}^{A-1}\frac{1}{2}\kappa(x_{j,\pm}-x_{j+1,\pm})^2+\frac{\gamma}{2}x_{n,\pm}^2\,,\nonumber\\
H_{S,\pm}&=&\frac{P^2_\pm}{2M} + \frac{1}{2}M\Omega_\gamma^ 2 X^2_\pm
\label{eq:Hspm}\,.
\end{eqnarray}
Here $\Omega_\gamma=\sqrt{\Omega^2+\gamma/M}$ is a shifted trap frequency and by definition $x_0=x_{0,+}/\sqrt{2}$, while $x_{0,-}=0$. One can note that the Hamiltonian terms are either symmetric (even parity) or antisymmetric (odd parity) and the sets of symmetric and antisymmetric coordinates therefore form two separate, uncoupled systems. 

If both defects couple to the same chain particle, i.e. $n=0$, the above equations show that only the defect COM coordinate couples to the symmetric coordinates of the chain, while the defect relative motion is perfectly decoupled and thus it is an eigenmode of the whole system at frequency $\Omega_{\gamma}$. Therefore,  under the conditions for which the chain acts as thermal bath for a single defect, it will induce thermalization of the COM mode of the defect particle and possible initial correlations between it and the relative motion of the defects are washed out. After a transient time the relative motion of the defects will therefore be in a state that is solely determined by the initially prepared state, while the COM will be in a thermal state at temperature $T$. This dynamic is the key element for entanglement generation between the oscillators. For instance, it can be shown that, if the relative motion is in a squeezed state and the temperature of the COM is sufficiently low, the product of the two orthogonal quadratures $\Delta X_-\Delta P_+$ (here taken in the reference frame rotating at the oscillator frequency $\Omega_\gamma$), can be below the standard quantum limit so that the two defects are two-mode squeezed, and thus entangled \cite{EPR,Reid,Duan:2000}. The squeezing of the relative motion can result from preparing each individual defect oscillator in a squeezed state at time $t=0$. This situation has been extensively analysed in Ref.~\cite{Kajari:2012} and is based on the existence of spatially-localized eigenmodes, which can be considered a realization of decoherence-free subspaces \cite{Lidar:1998}. 

\begin{figure}[t]
\begin{center}
\includegraphics[angle=0,width=\columnwidth]{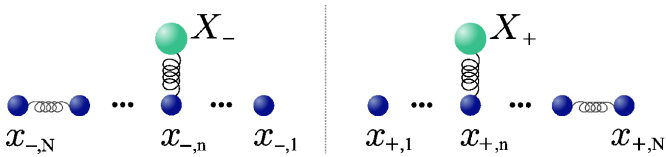}
\caption{(Color online) The COM and relative coordinates $X_{\pm}=(X_1 \pm X_2)/\sqrt{2}$ of the two defects are shown, which couple to two different reservoirs given by the coordinates $x_{i,\pm}=(x_i \pm x_{-i})/\sqrt{2}$.}
\label{fig:comrel}
\end{center}
\end{figure}

In the following we consider the situation in which the two oscillators are at a finite distance $d=2n$ for $N$ even and $d=2n-1$ for $N$ odd. This scenario is shown in Fig.~\ref{fig:comrel} where the COM and relative coordinates of the defects couple to two separate environments, composed of the chain of symmetric and antisymmetric displacements, respectively. For $n\neq 0$, when the impurity defects couple to different chain particles, there exists no eigenmode of the chain which solely involves displacements of the impurity defects. Nonetheless, the presence of the impurity defects breaks the discrete translational invariance of the chain, and gives rise to localized modes which under certain conditions can be eigenmodes of the chain. This can be verified by analysing the structure of the equations of motion. Consider for instance the equation of motion of oscillator $x_{n,+}$
\begin{equation}
\ddot{x}_{n,+}=-\left(\omega^2+\frac{\kappa}{m}\right)x_{n,+}-\frac{\gamma+\kappa}{m}(x_{n,+}-Q_+)+\frac{\kappa}{m}x_{n+1,+}\,,
\end{equation}
where $Q_+\propto(\gamma X_++\kappa x_{n-1,+})$. It is simple to show that one can construct a mode $Q^D_j$, which is a superposition of $X_+$ and $x_{0,+},x_{1,+},\ldots,x_{n-1,+}$ and which is orthogonal to $Q_+$ in the Riemannian space
with metric tensor ${\mathcal M}$, where  ${\mathcal M}$ is a diagonal matrix whose diagonal corresponds to $(M,m,\ldots,m)$ for the array $(X_+,x_{0,+},\ldots,x_{n,+})$ \cite{MorigiWalther:2002}. One can construct $n$ such modes. However, only one specific mode $Q^D_{j_0}$ can be an eigenmode of the dynamics. This happens when the frequency of the defect oscillator, $\Omega_\gamma$, matches a specific value $\omega_{j_0}^+$ \cite{Wolf:2011}. For instance, when $n=1$ one finds the localized mode $Q^D\propto( \kappa\sqrt{m/M}X_+-\gamma x_{1,+})$, which is an eigenmode of the composite dynamics when $\Omega_{\gamma}=\sqrt{\omega^2+\kappa/m}$. Analogous considerations can be made for the relative motion.

Having identified the localized modes, then the generation of entanglement between the two defect oscillators follows a similar route as the one described when the defects couple to the same chain particle. There are however two important differences. First, the projection of the defect oscillators into the decoupled mode is now smaller and therefore the initial squeezing only partly determines the variance of each composite quadrature. Moreover, the variance of each composite quadrature is also determined by the initial temperature of the chain, since there is a finite projection of the chain thermal state onto the decoupled mode (through the interposed chain oscillators). This limits the amount of entanglement one can reach when the oscillators are at a finite distance. In Sec. \ref{sec:dist_dependence} we will provide results for specific parameters.

\subsection{Discussion}
The stationary properties of the system above have been analysed in \cite{Wolf:2011,Kajari:2012}, and shown to be independent of the chain size and therefore applicable to the thermodynamic limit. The reason is that the localized modes involve just the defect oscillators and the interposed chain particles, and do not depend on the length of the chain as long as finite-size effects can be neglected. Entanglement in a chain of {\it identical} oscillators has also been investigated and in Refs.~\cite{Audenaert:2002,Anders:2008} the authors characterized the entanglement between two components at steady state. Studies on dynamical effects have highlighted the existence of entanglement between the ions at the chain edges \cite{Plenio:2004}, however this is not a stationary effect, as it vanishes in the thermodynamic limit for infinitely long chains.

Entanglement generation between two physical systems, such as spins or oscillators, has also been discussed by modeling the bath using the Born-Markov master equation in quantum-optical systems \cite{Huelga:2002,Benatti:2003}, or by resorting to phenomenological models, see for instance Refs. \cite{Braun:2002,Paz:2008,Zell:2009,Galve:2010}. In our model, one could consider to derive a master equation describing the dynamics of the two defects. Due to the presence of the localized mode, the bath of oscillators we consider is non-Markovian. A possible approach is to identify the localized eigenmode as a pseudomode coupling with the defects and thus analyze the dynamics using methods developed in Ref. \cite{Mazzola:2009}.  It is further interesting to consider measures of non-Markovianity for our system, as the ones proposed in Ref. \cite{Wolf:2008,Breuer:2009,Vasile:2011} and applied in similar settings in Ref. \cite{Liu:2011}. 
%Here, entanglement is not generated by a coherent action of the environment on the impurity oscillators, but rather by their non-Markovian dynamics. As discussed in Sec.~\ref{sec:dist_dependence}, the distance between which significant entanglement can be created greatly exceeds the thermal coherence length of the chain environment (typically on the order of its microscopic constituents \cite{Zell:2009}).

%%%%%%%%%%%%%%%%%%%%%%%%%%%%%%%%%%%%%%%%%%%%%%%
%%%%%%%%%%%%%%%%%%%%%%%%%%%%%%%%%%%%%%%%%%%%%%%
\section{Ion Coulomb chain with impurity defects}
\label{sec:model}

We now turn to investigate entanglement generation between two impurity defects embedded in a chain of trapped ions. With respect to the previous simplified model, the particles now interact via the long-range Coulomb repulsion. Moreover, they are generally confined in external potentials, which make the ions density inhomogeneous. A simple picture in terms of localized modes does not strictly apply, nevertheless we show that this model already provides a useful guidance to understand the dynamics in an ion chain, even for a small number (ten) of ions. 

Before we start, let us comment on previous work. In Ref. \cite{Serafini:2010}  entanglement transfer between the transverse modes of the ions was analysed. In Ref.  \cite{Cormick:2010} it was shown that stationary entanglement between two ions at opposite edges of a cluster composed of three aligned ions can be created. In this latter work the thermal bath was effectively provided by continuous sympathetic cooling of the central ion, and the effective energy transfer into the modes of the electromagnetic field gave rise to a dynamics analogous to thermalization. In the current work, instead, the bath is microscopically modeled by the lighter ions of the chain and this is what we discuss in the present section.

\subsection{An ion Coulomb chain in a linear Paul trap}

We consider $N$ ions with equal charge $Q$, which are confined in a linear Paul trap at positions ${\bf r_j}=(x_j,y_j,z_j)$. The trap secular potential reads
\begin{align}
V_{\rm trap}({\bf r_j})=(U_\parallel z_j^2+U_{\perp,j}(x_j^2+y_j^2))/2\,,
\end{align}
where $U_\parallel$ and $U_{\perp,j}$ determine the strength of the axial and transverse potentials. The latter is generated by a radio-frequency trap and depends on the ion mass $m_j$ via the relation $$U_\perp(m_j)=(U_0/m_j-U_\parallel)/2\,.$$  Here, $\sqrt{U_0}=Q\chi/(\sqrt{2}\Omega_{rf})$ with $\Omega_{rf}$ being the radio-field frequency and $\chi$ a constant depending on the trap geometry \cite{Kielpinski2000a}. Ions of different mass therefore experience different transverse trapping frequencies, and this mass dependence will be key for the desired dynamics.

When the ions are laser-cooled, they crystallize at the equilibrium positions of the total potential given by $V_{\rm trap}({\bf r})$ and the mutual Coulomb repulsion
\begin{align}
V=\sum_{j=1}^N V_{\rm trap}({\bf r_j})+\frac{Q^2}{8\pi\epsilon_0}\sum_{j\neq i}\frac{1}{|\br_i-\br_j|}.
\label{eq:full_potential}
\end{align}

\begin{figure}[tb]
\begin{center}
\subfigure[]{\includegraphics[width=0.49\columnwidth]{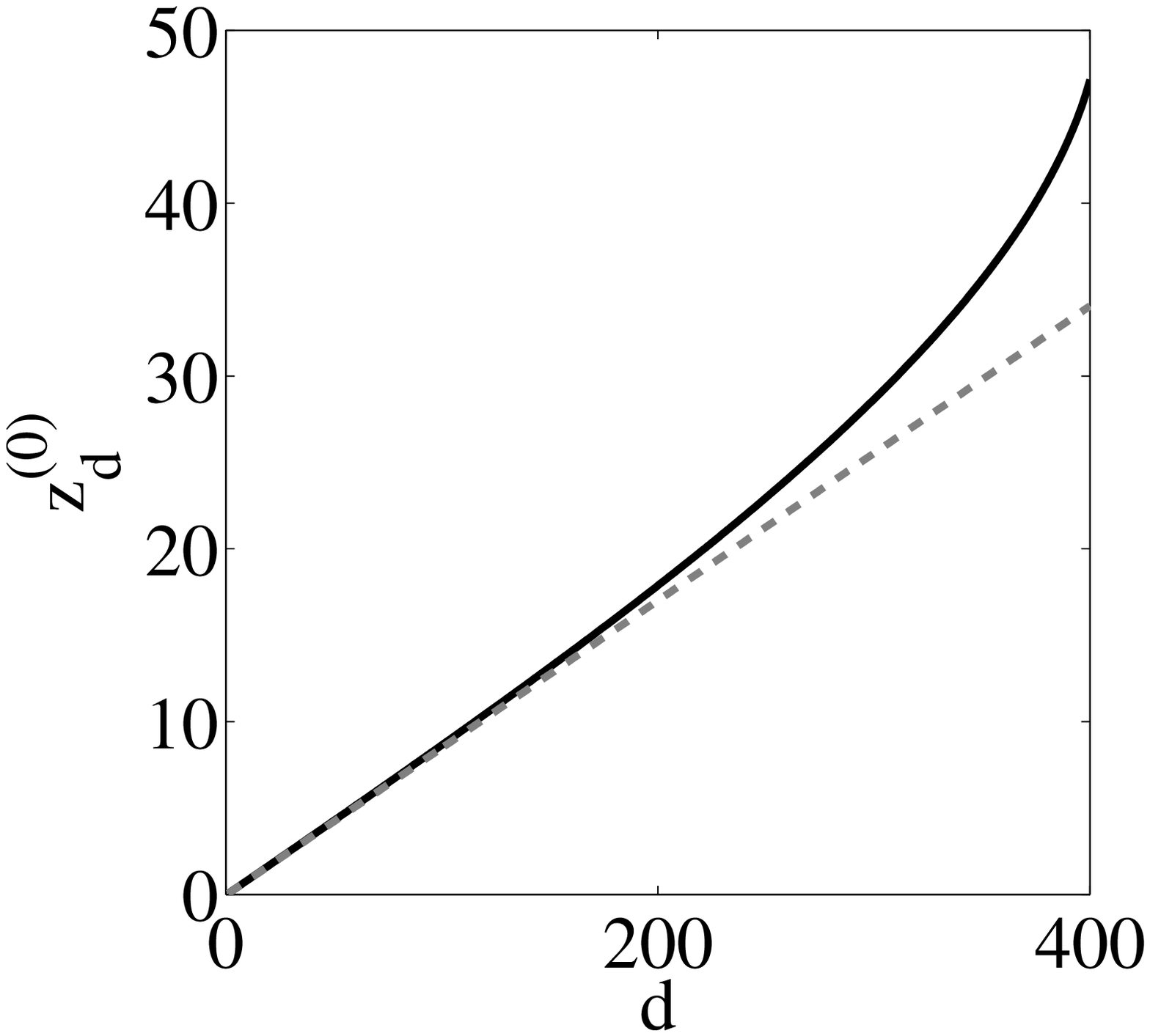}}
\subfigure[]{\includegraphics[width=0.49\columnwidth]{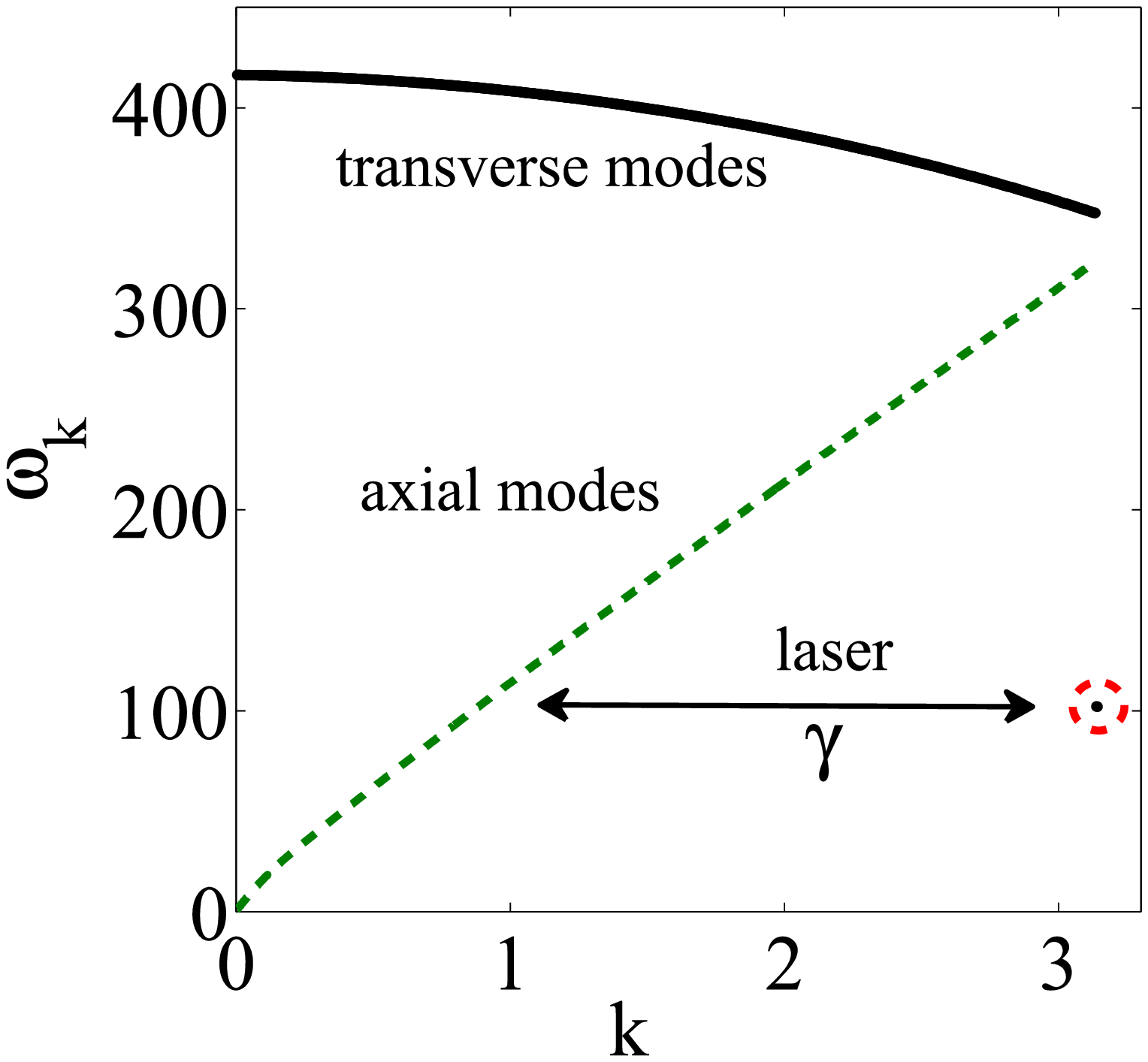}}
%\label{fig:spectrum}}
\caption{(Color online) (a) Equilibrium distance $z_d^{(0)}=z_n^{(0)}-z_{-n}^{(0)}$ between symmetrically placed pairs of ions in a linear Paul trap as a function of the ions dimensionless relative position $d=2n$ (solid line). The distance $z_d$ is in units of the interparticle distance $a$ at the chain center and has been evaluated for $N=400$ ions. The dashed line represents the distance when the ion distribution is uniform and equally spaced by the interparticle distance $a$. Around the trap centre the spacing in a linear Paul trap is well approximated by constant ion spacing (and is fitted by $\approx2.29 N^{-0.596}$ \cite{Steane:1998}). (b) Axial and transverse spectrum of a chain composed of Ca$^+$ ions in a linear Paul trap, which has two In$^+$ ions embedded in it. The distance between the Calcium ions is $d=14$ and the mass ratio in this system is $M/m =2.87$ \cite{Hayasaka2012a}. The solid (dashed) line corresponds to the eigenfrequencies $\omega_k$ of $H_\perp$ ($H_\parallel$) and is plotted as a function of the quasimomentum $k$ (in units of $\pi/a$). The frequencies are scaled in units of $\omega_\parallel=\sqrt{U_\parallel/m}$ and $U_\perp(m)/U_\parallel\simeq416$ for any $N$ according to Eq.\eqref{eq:trapscale}. The isolated transverse frequencies (see dashed circle) almost coincide with the frequency of the defects transverse potential.}
\label{fig:Dist}
\end{center}
\end{figure}

Here we choose the trap aspect ratio $\epsilon=U_\perp(m)/U_\parallel$ to be sufficiently large, so that the ions crystallize along the $z$ axis at the equilibrium positions $z_j^{(0)} $, which are the solutions to
\begin{equation}
U_\parallel z_j^{(0)} + \frac{Q^2}{4\pi\epsilon_0}\left( \sum_{i=1}^{j-1}\frac{1}{(z_i^{(0)}-z_j^{(0)})^2}-\sum_{i=j+1}^{N}\frac{1}{(z_i^{(0)}-z_j^{(0)})^2}\right)=0.
\label{eq:equilibrium_positions}
\end{equation}
Note that in the longitudinal direction none of the parameters depend on the ions mass and that the resulting interparticle distance at equilibrium is not uniform \cite{Dubin:1997,Steane:1998}.  Therefore, to ensure that the Hamiltonian is symmetric with respect to mirror reflection about the chain centre, which is a prerequisite for the existence of localized eigenmodes involving the defects oscillations, the defect modes have to be placed symmetrically with respect to the chain centre \cite{Morigi:2004}. In Fig.~\ref{fig:Dist}(a) we show how the distance between pairs of defect ions scales as a function of the number of interposed ions. While for a large system a certain robustness with respect to the exact positioning of the defects can be expected, in a small system the defects have to be precisely located. This requirement is relaxed in settings in which the ions are axially equidistantly spaced, for example in a chain of ions at the central axis of a three dimensional crystal \cite{Drewsen:centralchain} or when the ions are confined by means of anharmonic potentials \cite{Duan,Champenois:2010}. 
 
\subsection{Coupled oscillators}

\begin{figure}[h]
\begin{center}
\subfigure[]{\includegraphics[width=0.49\columnwidth]{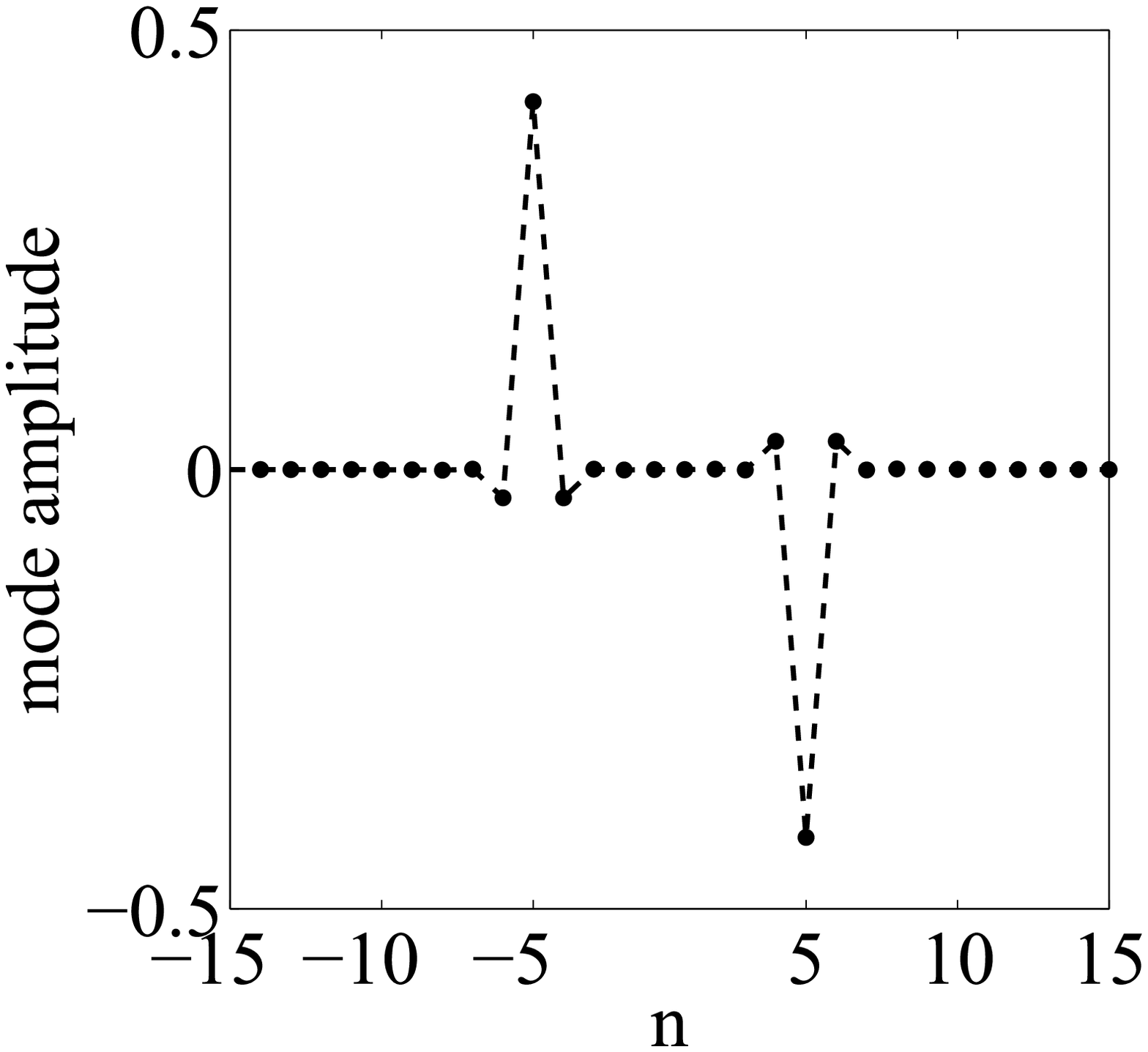}}
\subfigure[]{\includegraphics[width=0.49\columnwidth]{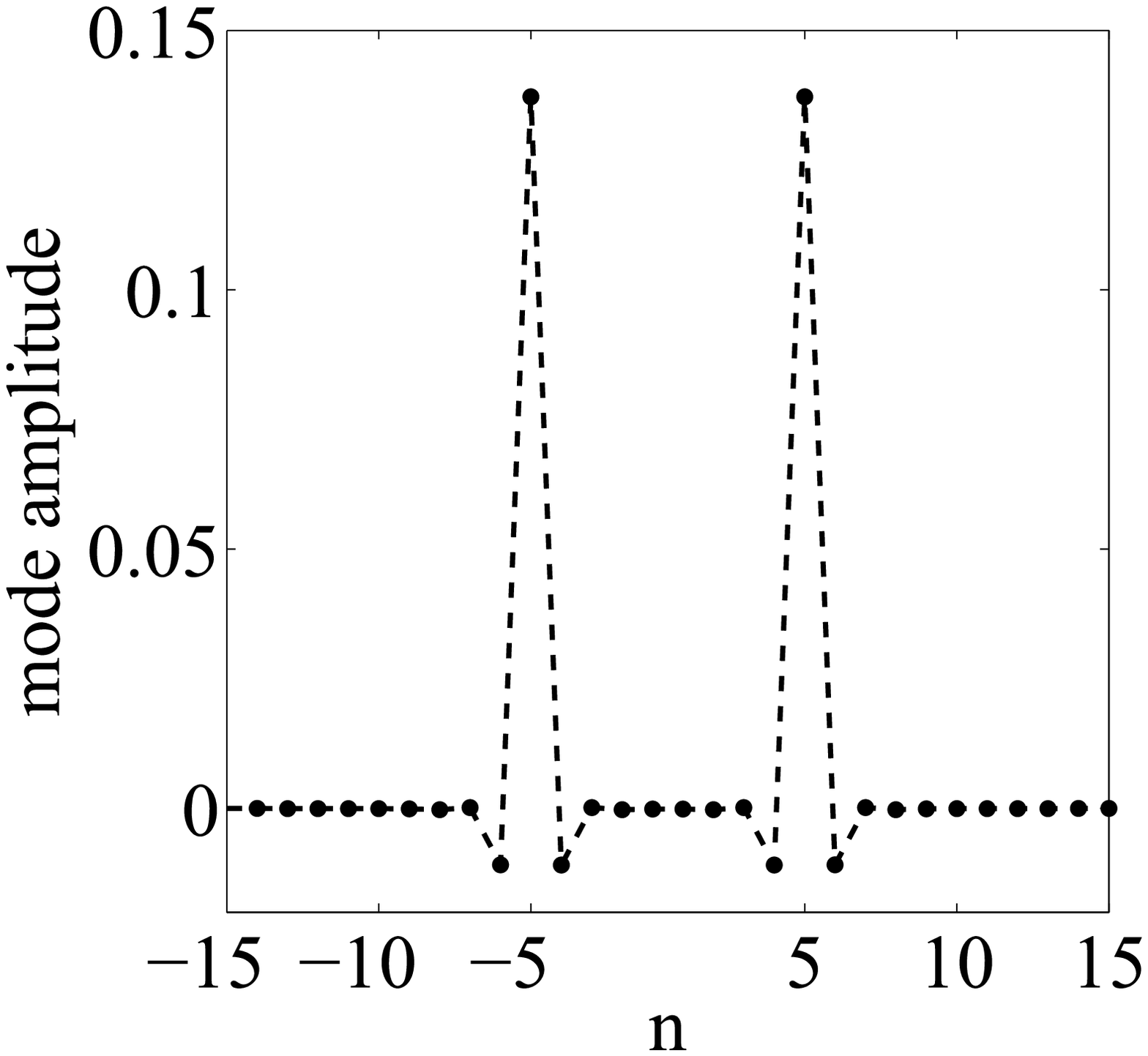}}
\caption{Spatial distribution of the isolated modes  along the ion chain with dimensionless mode number $n$ for two defects separated by $d=8$ in a chain of size $N=400$. The figure shows (a) the anti-symmetric relative positions mode and (b) the symmetric COM mode. Apart for the defects positions, only the first neighbours contribute significantly to the mode dynamics. Ions outside the plotted region also possess no overlap with the isolated modes.}
\label{fig:eigenmode_imp}
\end{center}
\end{figure}

We now assume that the chain has been prepared at a temperature $T$ such that the ions carry out harmonic oscillations around their equilibrium positions $\br_j^{(0)}=(0,0,z_j^{(0)})$. The potential can hence be described by its harmonic approximation, and under this condition, axial and transverse dynamics decouple. The Hamiltonian can then be written as 
\begin{equation}
\label{H:0}
H_0=H_{\parallel}+H_{\perp}^{(x)}+H_{\perp}^{(y)}\,, 
\end{equation}
with

\begin{align}
\label{H:parallel}
&H_\parallel=\sum_{j=1}^N\left(\frac{p_{q,j}^2}{2m_j}
+\frac 12\sum_{k=1}^NV_{jk}^\parallel\,q_j\,q_k \right),\\
&H_\perp^{(x)}=\sum_{j=1}^N\left(\frac{p_{x,j}^2}{2m_j}
+\frac 12\sum_{k=1}^NV_{jk}^\perp\,x_j\,x_k \right).
\label{eq:hamiltonian1}
\end{align}
Here $q_j=z_j-z_j^{(0)}$ and $p_{j,\alpha}$ are the ${\alpha=x,y,q}$ components of the momentum $\bp_{j}$ of ion $j$.  The components of the axial and transverse potentials read, respectively,
\begin{align}
&V_{jk}^\parallel = \left(U_\parallel + \sum_{l\ne j}\mathcal K_{jl}\right)\delta_{jk} - (1-\delta_{jk})K_{jk},\\
&V_{jk}^\perp = \left(U_{\perp,j} - \frac 12 \sum_{l\ne j}\mathcal K_{jl}\right)\delta_{jk} + (1-\delta_{jk})\frac{K_{jk}}2,
\end{align}
where $\delta_{jk}$ is the Kronecker delta, while $\mathcal K_{j,\ell}=2Q^2/(4\pi\epsilon_0\,|z_j^{(0)}-z_\ell^{(0)}|^3)$ are the couplings due to the Coulomb repulsion \cite{Morigi:2004}. The Hamiltonian term $H_{\perp}^{(y)}$ is found from $H_{\perp}^{(x)}$ by replacing $x_j\to y_j$. %The form of Hamiltonian $H_0$, Eq. \eqref{H:0}, shows that axial and transverse vibrations of an ion chain are decoupled in the harmonic limit. 

In Fig.~\ref{fig:Dist}(b) we show the eigenfrequency spectrum for a chain composed of $N-2$ Ca$^+$  ions, into which two In$^+$ ions  are embedded such that $14$ Calcium ions are interposed. This system has a mass ratio of $M/m\approx 2.87$ \cite{Hayasaka2012a} and in order to be able to easily compare results for chains of different length we rescale $\omega_{\parallel}=\sqrt{U_\parallel/m}$ so that \cite{Morigi:2004}
\begin{equation}
\omega_{\parallel}(N)=\omega_\text{ref}\frac{\log N}{N}\, ,
\label{eq:trapscale}
\end{equation}
where $\omega_\text{ref}=2\pi\times 659.6$ kHz is a reference axial trap frequency, which we have chosen such that the interparticle distance at the centre of the chain is constant for all $N$ and therefore the dispersion relation remains invariant as $N$ is varied. The spectrum consists of two degenerate normal frequencies of the transverse modes (TM), which appear separate from the continuum (four isolated frequencies in total, two for each transverse spectra). They correspond to normal modes localized around the position of the two defects and occur due to the mass dependence of the transverse, radio-frequency potential. By varying the mass ratio $\mu=M/m$ or the trap aspect ratio $\epsilon$, these localized frequencies can be tuned and for large enough separation of the continuum part from the TM spectrum, the dynamics of these two modes decouple from the rest of the transverse chain. The largest amplitudes are observed at the position of the defect ions and their first neighbours (see Fig.~\ref{fig:eigenmode_imp}), while all other ions are essentially unaffected. We note that these degenerate modes are either symmetric or anti-symmetric with respect to reflection about the chain centre.

In order to realize a dynamics analogous to the one sketched in the previous section, we need to introduce a coupling between the transverse defect modes and the axial ones. Furthermore, the frequency of the radial defect modes needs to fall into the same range as the band of the axial excitations (see Fig.~\ref{fig:Dist}(b)). The second condition can be easily achieved by appropriately choosing the mass ratio and the trap confinement. Moreover, while anharmonicities (i.e., higher order terms of the Taylor expansion of the Coulomb interaction) can give rise to a non-negligible coupling between the radial defect vibration and the axial vibrations, for sufficiently cold atoms these only become important over time scales longer than the revival time of the Gaussian dynamics. A coupling like the one given in Eq.~\eqref{eq:HIF}, however, can originate from the interaction with a standing-wave laser field in the $x-z$ plane, which is switched on at time $t=0$ and dispersively couples with an internal transition of the impurity defect ions. When the node of the laser standing wave coincides with the equilibrium positions of the defect ions, the chain dynamics in the Lamb-Dicke regime \cite{Cirac1992} is governed by the Hamiltonian $$H=H_0+H_I(t)\,,$$ where
\begin{equation}
  H_I(t)=\frac{\gamma(t)}{2}\left[(x_{-n}-q_{-n})^2+(x_{n}-q_{n})^2\right]\,,
\end{equation}
with $\gamma(t)=\gamma\,\Theta(t)$ being an effective coupling strength and $\Theta(t)$ the Heaviside function. This local coupling allows an excitation in the defects' TMs to also excite their axial degree of freedom, which in turn generates a phononic excitation in the axial direction. Note that the dynamics in the $y$-direction is decoupled and therefore ignored in the following.

\subsection{Initial state and Gaussian dynamics}

With this picture in mind, our goal will be to entangle the two defects TMs through their interaction with a reservoir provided by the axial phonons. The initial states of the axial modes of the chain are prepared by Doppler cooling, so that the density matrix for the axial oscillators reads $\rho_R(T)=\exp(-H_R/k_BT)/Z$, where $k_B$ is the Boltzmann constant, $T$ is the temperature characterising the dynamical steady state obtained by laser cooling \cite{Morigi:2001}, $H_R\equiv H_{\|}$ is the reservoir Hamiltonian and $Z=\textmd{Tr}[\exp(-H_R/k_BT)]$ its partition function. The TMs of the defect are initially prepared in the ground state of the transverse oscillator through sideband cooling \cite{Leibfried:2003} and then converted into squeezed pure states \cite{Wineland:NIST,Leibfried:2003}, described by the density matrix $\rho^{(1)}_{n}(s)\otimes\rho^{(2)}_{-n}(s)$, with a real-valued squeezing parameter $s$ and variances $\Delta q_{n}^2=\Delta q_{-n}^2=x_0^2 e^{-2s}/2$, $\Delta p_{q,n}^2=\Delta p_{q,-n}^2=p_0^2 e^{2s}/2$. Here $x_0=\sqrt{\hbar/M\omega_{\perp}}$ is the size of the defect ground state in the transverse direction with frequency $\omega_\perp=\sqrt{U_\perp(M)/M}$ and $p_0=\hbar/x_0$ is the associated momentum \cite{Braunstein2005}. The initial state of the composite system is then
\begin{align}
\varrho(0)=\rho^{(1)}_{n}(s)\otimes\rho^{(2)}_{-n}(s)\otimes\rho_R(T)\,.
\label{eq:initial_state}
\end{align}
Under the prescribed Hamiltonian, $H_0$, the states of the defects (as well as the reservoirs) remain Gaussian and are therefore fully characterized by their first moments and covariance matrix $\Sigma_{ij}=\frac12\left<\xi_i\xi_j+\xi_j\xi_i\right>-\media{\xi_i}\media{\xi_j}$, with $i,j\in\{1,2,3,4\}$ and $\xi=\left(q_{j_1},p_{j_1,q},q_{j_2},p_{j_2,q}\right)$ \cite{Adesso2007}. Logarithmic negativity can thus be used to quantify entanglement between the defects, according to 
\begin{equation}
E_N=\max\{0,-\ln(2\tilde\nu_-)\}\, ,
\label{eq:logneg}
\end{equation}
where $\tilde\nu_-$ is the smallest symplectic eigenvalue of the partial transpose of the covariance matrix $\Sigma$ \cite{Vidal:2002}. 

\subsection{Mirror reflection symmetry}

Let us consider that the defects are placed symmetrically with respect to the trap centre, so that we can use COM and relative coordinates for pairs of ions 
\begin{align}
\label{Trafo}
&q_{j,\pm}=(q_{j}\pm q_{-j})/\sqrt{2},\\
&x_{j,\pm}=(x_{j}\pm x_{-j})/\sqrt{2},
\label{Trafo:1}
\end{align}
where the index $+$ ($-$) indicate COM (relative) motion and their conjugate momenta are defined accordingly. As before, the full Hamiltonian decouples in these new variables and can then be written as $H=H^++H^-$ with
\begin{align}
H^\pm=H_\parallel^\pm+H_\perp^\pm+H_I^\pm\,,
\end{align}
and where $H_I^+$ and $H_I^-$ are the coupling terms between the axial and transverse directions. The non-local dynamics between the defects can therefore be described by two independent couplings to individual environments. In the presence of the coupling laser the axial potential, $(V_\parallel^{(\gamma)})_{jk}=(V_\parallel)_{jk}+\gamma\delta_{jk}(\delta_{j,n}+\delta_{j,-n})$, still decouples into a COM and a relative part, $V_{\parallel,\pm}^{(\gamma)}$, and by introducing the mass-weighted coordinates $q_{j,\pm}^\prime=\sqrt{m_j}q_{j,\pm}$, the corresponding potential $(V_\parallel^{(\gamma)})_{jk}^\prime=(V_\parallel^{(\gamma)})_{jk}/\sqrt{m_jm_k}$  can be diagonalized by means of an orthogonal matrix \cite{MorigiWalther:2002}. The eigenvalue problem is now equivalent to the one of $N$ identical ions of unit mass
\begin{align}
H_{\parallel,\pm}^{(\gamma)} = \sum_{j=1}^N\left(\frac{[\rho_{q,\pm}]_j^2}{2} + \frac{(\omega_{\pm,j}^{(\gamma)})^2[\chi_\pm]_j^2}{2}\right),
\label{eq:wpm}
\end{align}
where $(\omega_{\pm,1}^{(\gamma)},\cdots,\omega_{\pm,N}^{(\gamma)})$ denote the eigenfrequencies of $V_{\parallel,\pm}^{(\gamma)\prime}$ and $O_{\parallel,\pm}$ is the orthogonal matrix which brings $V_{\parallel,\pm}^{(\gamma)\prime}$ into a diagonal form. The eigen-positions and momenta are given by $[\chi_{\pm}]_j=\sum_k[O^T_{\parallel,\pm}]_{jk}[q_\pm^\prime]_k$ and $[\rho_{q,\pm}]_k=\sum_k[O^T_{\parallel,\pm}]_{jk}[p_{q,\pm}^\prime]_k$. The transverse coordinate is equivalently diagonalized.
%%%%%%%%%%%%%%%%%%%%%%%%%%%%%%%%%%%%%%%%%%%%%%%
%%%%%%%%%%%%%%%%%%%%%%%%%%%%%%%%%%%%%%%%%%%%%%%

\subsection{Spectral density}

Some insight into the dynamics can be gained by using the generalized quantum Langevin equations of motions for $X_{\pm}\equiv x_{\pm,n}$, which can be obtained by formally eliminating the other variables \cite{Weiss:1999}
\begin{align}
\label{HLE}
\frac{d^2X_\pm}{dt^2}&+\int_0^t dt' \Gamma_\pm(t-t') \frac{dx_{\pm,n}}{dt'}+(1-\Gamma_\pm(0))X_\pm(t)
\nonumber\\
&=F_\pm(t)-\Gamma_\pm(t)X_\pm(0)\,.
\end{align}
Here, $\Gamma_\pm(t)$ is the memory-friction kernel for the symmetric ($+$) and antisymmetric ($-$) modes for $t\ge0$ (it vanishes for $t<0$), which is the sum of the memory-friction kernels due to the coupling with the axial and transverse modes, 
\begin{align}
\Gamma_\pm(t)&=\Gamma_{\|,\pm}(t)+\Gamma_{\perp,\pm}(t)\;.
\end{align}

Without a loss of generality, we will focus  for the present discussion on the properties of the reservoir constituted by the axial modes. This assumption is justified by the observation that the defect modes are within the axial band, while there is a gap separating the defect oscillators from the other transverse excitations (Note that when reporting numerical results we include {\it all} couplings). Also for brevity we will just discuss the COM case in the following, as the formalism is equivalent for the relative coordinate.

\begin{figure}[t]
\begin{center}
\includegraphics[width=0.4\textwidth]{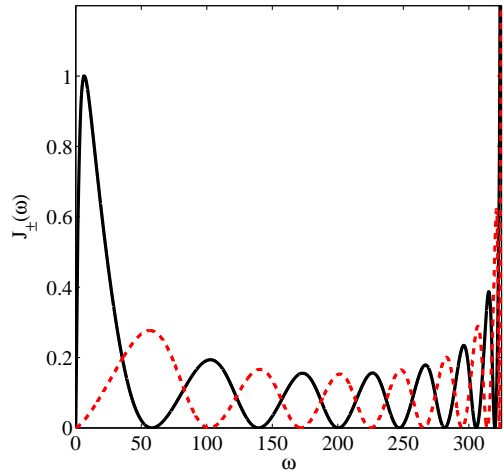}
\caption{(Color online)  Spectral densities (in units of $U_\parallel$) as a function of the frequency. The defects are separated by $d=14$ ions of the other species and the number of ions composing the chain is $N>d$. The solid (dashed) line corresponds to $J_+(\omega)$ ($J_-(\omega)$).} 
\label{fig:spec_density}
\end{center}
\end{figure}

The memory-friction kernel in the axial direction for the COM mode is
\begin{align}
\Gamma_{\parallel,+}(t)&=\sum_{j=1}^N\frac{(\gamma_{+,j})^2}{m(\omega^{(\gamma)}_{+,j})^2}\cos(\omega^{(\gamma)}_{+,j} t)\,,
\end{align}
where the new coupling strengths are defined from the eigenvectors of Eq.~\eqref{eq:wpm} as $\gamma_{+,j}=\gamma[O^T_{\parallel,+}]_{jn}$ (for defect ions at positions $n$ and $-n$). The operator $F_+(t)=F_{\|,+}(t)+F_{\perp,+}(t)$ is similarly decomposed into the contribution of the Langevin force due to the axial and the transversal modes, respectively. In particular,
\begin{align}
F_{\|,+}(t)=&\sum_{j=1}^N
\bigg[\gamma_{+,j}\cos(\omega^{(\gamma)}_{+,j} t)[\chi_{+}(0)]_j\notag\\
&+\frac{\gamma_{+,j}\sin(\omega^{(\gamma)}_{+,j} t)}{m\omega^{(\gamma)}_{+,j}}[\rho_{q,+}(0)]_j \bigg]\,.
\end{align}
The influence of the interaction with the reservoir can be characterised by the environmental spectral density, which is given by the Fourier cosine-transform of the memory-friction kernel
\begin{align}
J_+(\omega)= &\omega\int_0^\infty \Gamma_{\parallel,+}( t)\cos(\omega t)dt\nonumber\\
=&\frac{\pi}{2} \sum_{j=0}^A \frac{( \gamma_{+,j})^2}{ m( \omega_{+,j})^2}\,\delta(\omega- \omega_{+,j})\,,
\label{eq:spec_density}
\end{align}
where  $J_+(\omega)$ and the equivalent equation for $J_-(\omega)$ account for the distinct interactions of the centre-of-mass and relative defects TMs with the axial modes of the chain. For two defects separated by $d=14$ these spectral densities are shown in Fig.~\ref{fig:spec_density}  and the clearly visible zeros are a  manifestation of the existence of eigenmodes which are localized excitations within the chain. This means that at the frequency values $\omega^{(\ell)}_+$ [$\omega^{(\ell)}_-$] associated with these zeros, the COM [relative] motion of the defects effectively decouples from the environment, analogous to the effect described in Sec. \ref{Sec:MicroModel}.  

\subsection{Relevant time scales}

\begin{figure}[t]
\begin{center}
\subfigure[]{\includegraphics[height=0.457\columnwidth]{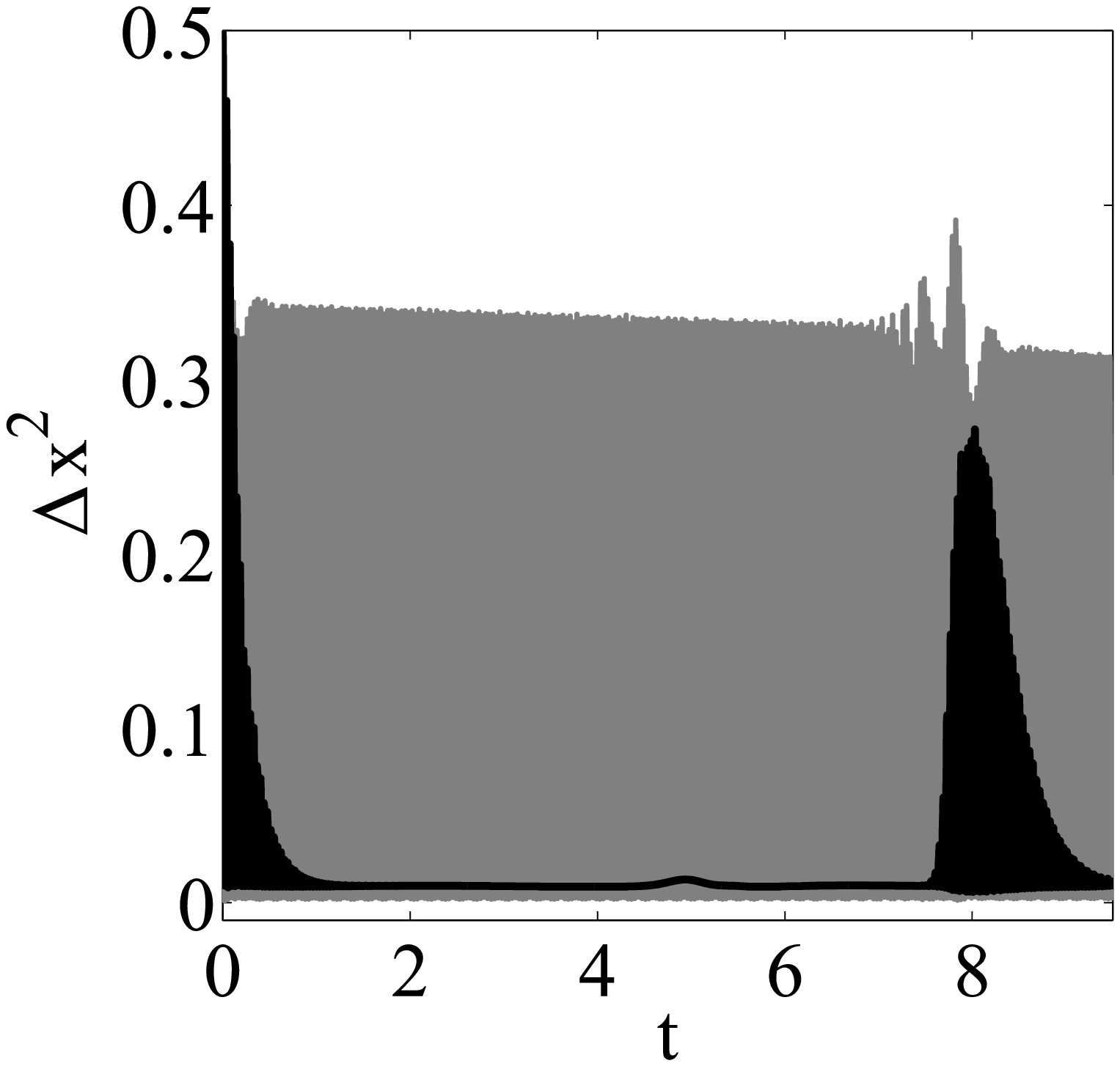}}
\subfigure[]{\includegraphics[height=0.45\columnwidth]{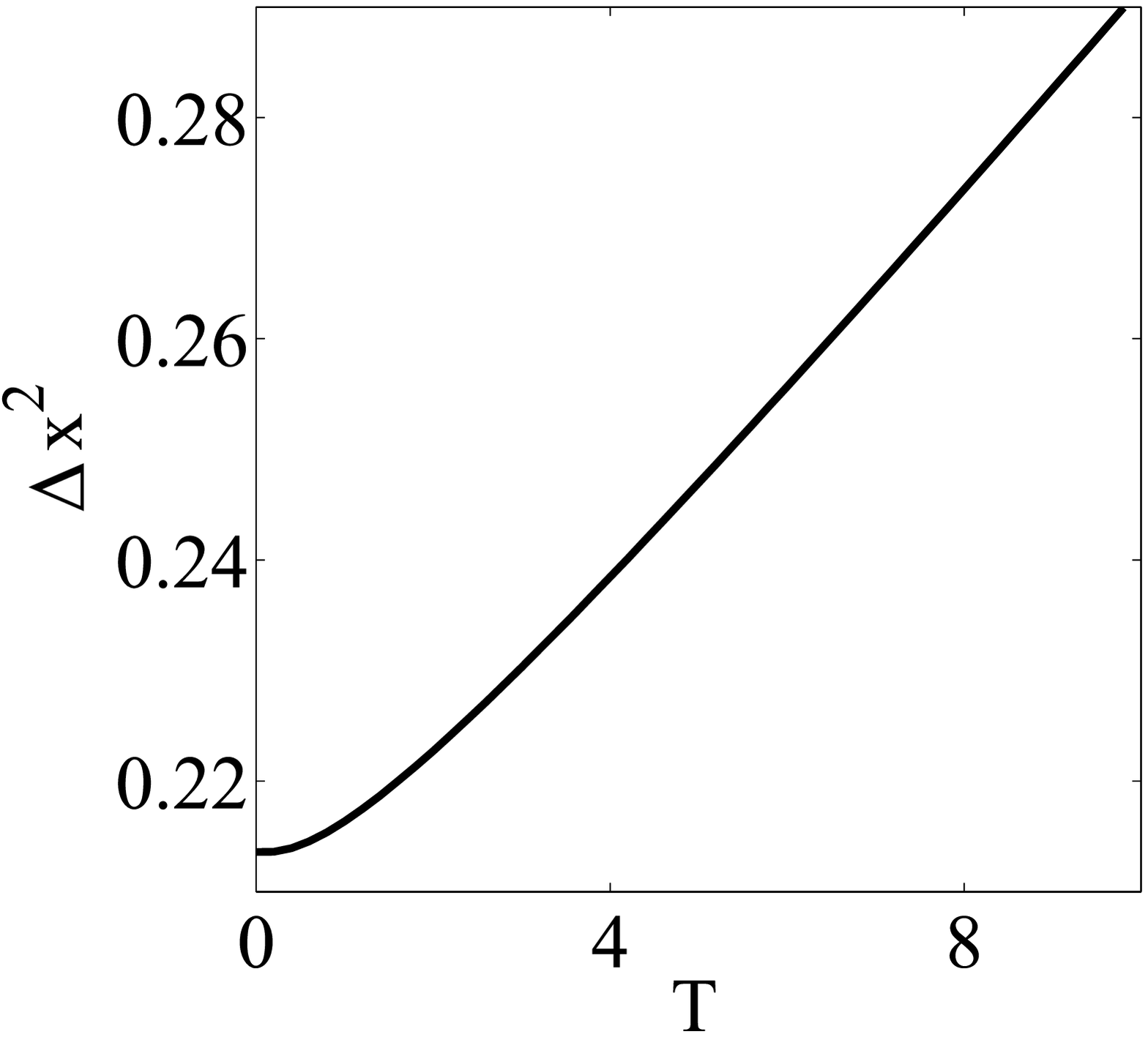}}
\label{fig:Deltax1}
\caption{(a) Expectation values of $\Delta x^2$ for the relative (grey line) and COM (black line) motions as a function of time (in units of $\omega_{\parallel}^{-1}$). The parameters are  $N=800$, $\gamma= 1.9 \times 10^3 U_\parallel$, $d=18$, $T=\hbar \omega_{\perp}/k_B$ and the defects frequency is set to $\omega^{(2)}_-$, corresponding to a zero of the spectral density. (b) Expectation value of $\Delta x^2$ for the COM motion at the steady state as a function of the initial temperature of the chain $T$ (in units of  $\frac{k_B}{\hbar \omega_{\parallel}}$).} 
\label{fig:transverse_moments}
\end{center}
\end{figure}

In the following we will analyze the defects dynamics by integrating Eqs.~\eqref{HLE} for the case where the frequency of the defect oscillators is such that the axial spectral density vanishes for $X^-$. We are in particular interested in the behaviour of the expectation values of the first and second moments of $X^+$ and $X^-$ and in Fig.~\ref{fig:transverse_moments} we show their evolution in time for a certain parameter choice. The defects frequency is chosen to match that of the second zero,  $\omega^{(2)}_-$, of the relative spectral density $J_-(\omega)$, and for times shorter than the revival time, the COM motion can be seen to reach a thermal quasi-steady state with zero mean transverse position and finite variance determined by the temperature $T$.
% and obeys equations analog to Eqs. (\ref{eq:sigma_qq}) and (\ref{eq:sigma_pp}).
%From this figure the COM defect is seen to thermalize with the reservoir. 

Figure~\ref{fig:transverse_moments}(a) allows us to identify two characteristic times scales: (i) the thermalisation time $t_\text{th}$, representing the time needed for the moments of the COM defect to reach quasi-steady values and (ii) the revival time $t_\text{rev}$, at which finite-size effects take this defect out of thermal equilibrium. Tuning the parameters to ensure a regime for which $t_\text{th}\ll t_\text{rev}$ then allows one to make statements which are valid in the thermodynamic limit. %Later on we will also check the behaviour for very small ion chains, which are experimentally accessible. 
In Fig.~\ref{fig:transverse_moments}(b) the average of the expectation value of $\Delta x^2$ is shown after thermalisation as a function of the initial temperature of the bulk and a linear increase for higher temperatures is visible. The hyperbolic cotangent dependence of $\Delta x^2$ on $T$, evident in Fig.~\ref{fig:transverse_moments}(b), confirms the expected thermalisation of the COM motion \cite{Kajari:2012}. We note that while the (non-local) COM/relative variables may reach a thermal state, this will not be true for the defect ions themselves as their dynamics also depends on the uncoupled, protected mode.
%%%%%%%%%%%%%%%%%%%

%%%%%%%%%%%%%%%%%%%%%%%%%%%%%%%%%%%%%%%%%%%%%%%
%%%%%%%%%%%%%%%%%%%%%%%%%%%%%%%%%%%%%%%%%%%%%%%
\section{Results}
\label{sec:Results}
%%%%%%%%%%%%%%%%%%%

To characterize the dynamics of the quantum correlations between the defects, we will focus first on the behaviour in long chains in order to understand scaling with the size of the bulk. We will then consider shorter chains, composed of tens of ions, and show that quantum state preparation -such as squeezing of the transverse modes of the defects and laser cooling of the rest of the chain- are the only relevant conditions for creating entanglement between the impurity ions. 

Since the initial defect state is Gaussian and the Hamiltonian quadratic, we will use the logarithmic negativity $E_N$, Eq.~\eqref{eq:logneg}, to quantify the quantum correlations between the defects. We will also present results related to the average logarithmic negativity $\overline{E_N}$, which is defined as
\begin{equation}
\overline{E_N}=\frac{1}{n}\sum_{i=1}^{n}E_N(\bar{t}_i) \, ,
\end{equation}
where $n$ is the number of time steps calculated in the interval $t_\text{th}<\bar{t}<t_\text{rev}$. Convergence of this mean value was tested rigorously as a function of $n$ over this interval.

%%%%%%%%%%%%%%%%%%%%%%%%%%%%%%%%%%%%%%%%%%%%%%%
%%%%%%%%%%%%%%%%%%%%%%%%%%%%%%%%%%%%%%%%%%%%%%%
\subsection{A reservoir can entangle two distant defects}
\label{sec:ent_generation}

As a benchmark, we start by analysing the entanglement generation when there is no coupling between axial and transverse modes, {\it i.e.} $\gamma=0$ and for time scales on which anharmonicities can be neglected. In this limit the evolution of the transverse modes of the defects is determined by the Hamiltonian $H_\perp$, Eq. \eqref{eq:hamiltonian1},  alone and we note that the two defect modes are directly coupled through the fluctuations of the Coulomb interaction about the equilibrium positions, which scale with their mutual distance $R_{12}$ as $1/R_{12}^3$. 

\begin{figure*}[t]
\begin{center}
\subfigure[]{\includegraphics[width=0.3\textwidth]{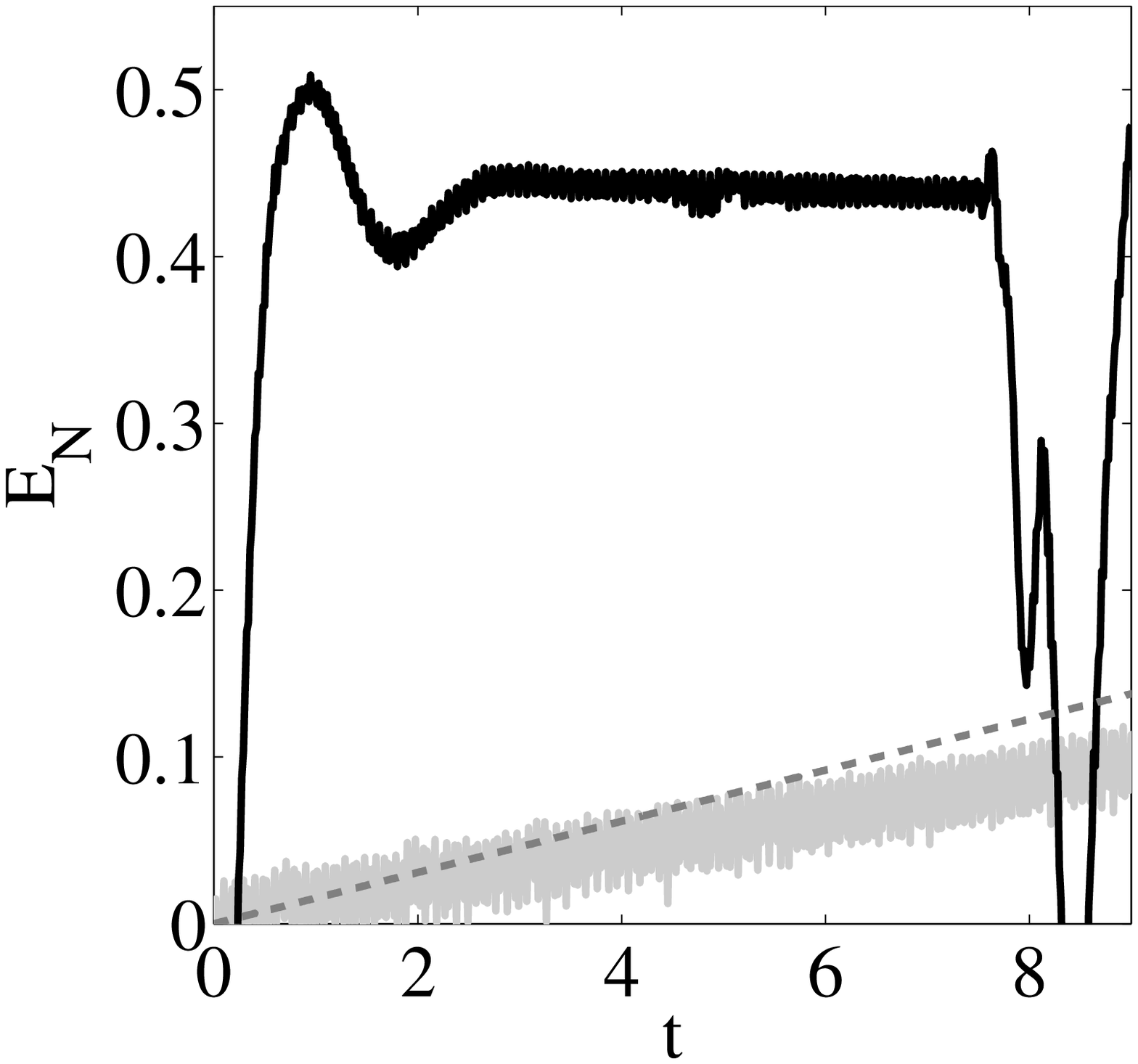}}
\subfigure[]{\includegraphics[width=0.3\textwidth]{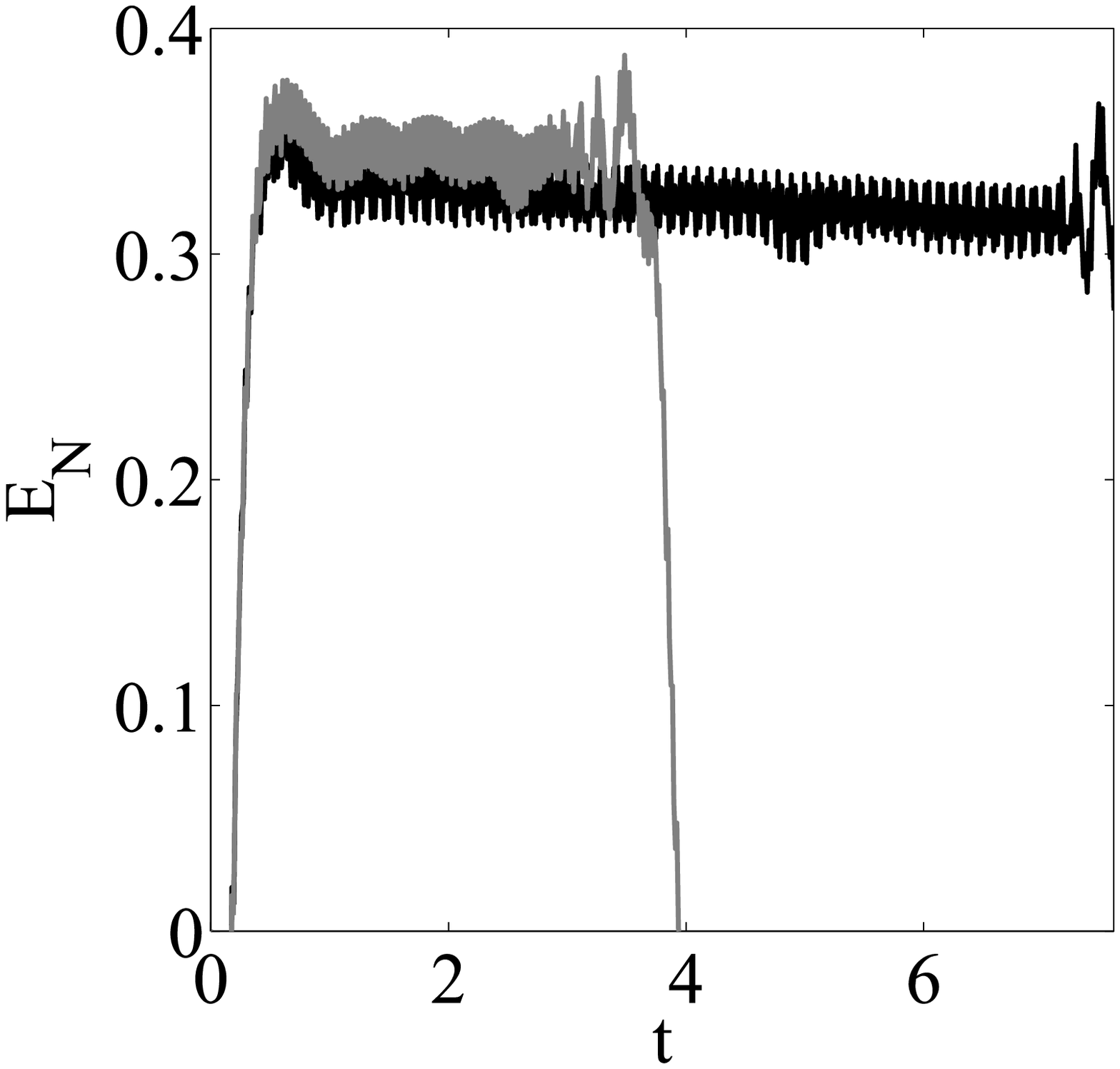}}
\subfigure[]{\includegraphics[width=0.3\textwidth]{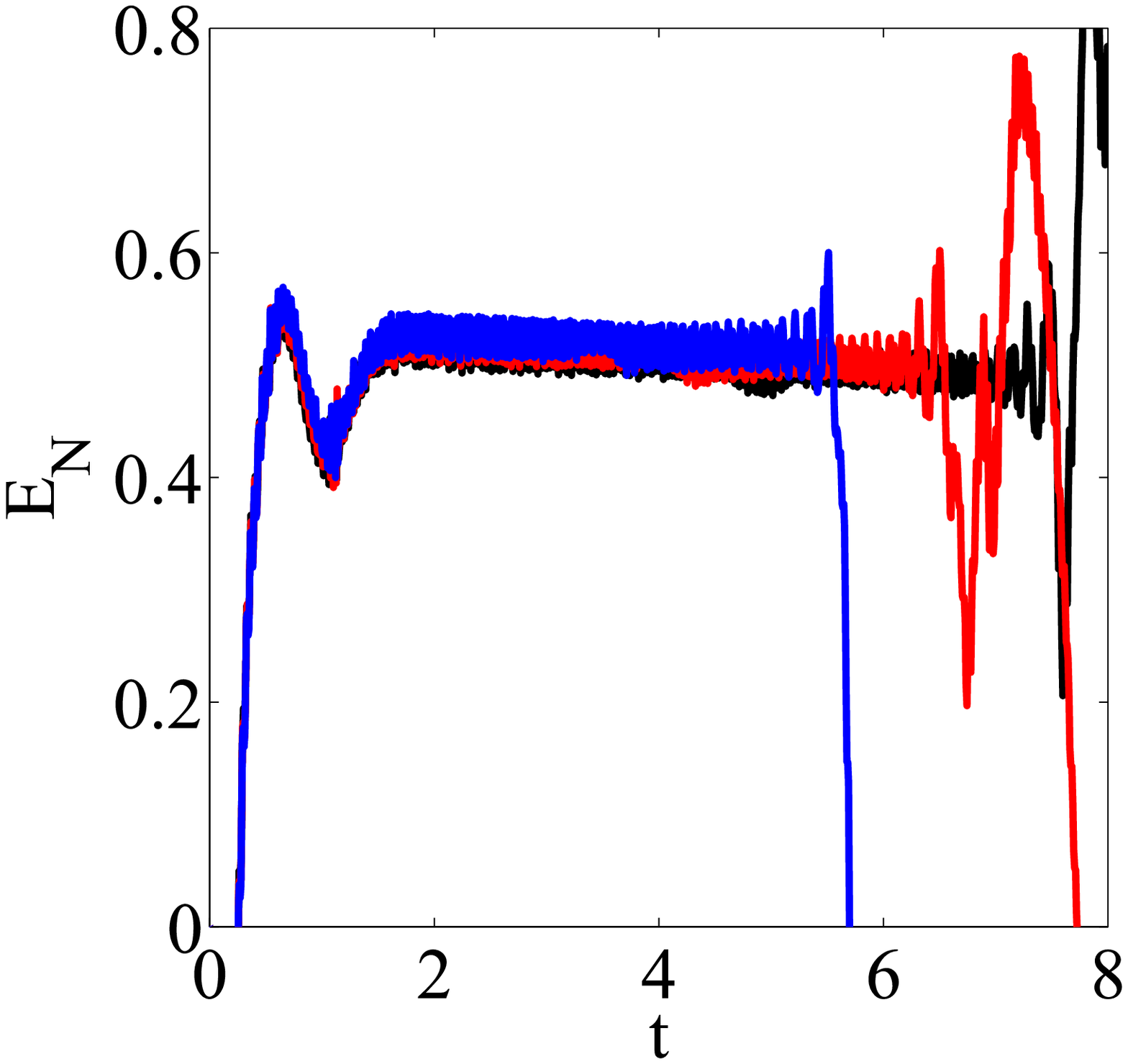}}
\caption{(Color online) (a) Logarithmic negativity as a function of time (in units of $\omega_{\parallel}^{-1}$) for a chain of $N=800$ ions and $d=18$ for an initial squeezing of $s=1.5$. The solid black line displays $E_N(t)$ in the presence of laser coupling with strength $\gamma= 1.9 \times 10^3 U_\parallel$ . For comparison the logarithmic negativity with no laser coupling ($\gamma=0$) is shown (light grey line) where the sole contribution of the transverse chain is seen to add small amplitude oscillations. Also shown is the entanglement generated between two isolated defects at the same distance (dashed grey line). (b) Comparison between the logarithmic negativity generated using the equilibrium positions determined by Eq.(\ref{eq:equilibrium_positions}) (black line) and by assuming equally spaced ions such that $a_n=z_{n+1}-z_{n}$ is constant (dark grey line). In both cases the distance between the defects is $d=18$ and $N=800$ with $\gamma=3.5 U_{\parallel}$. The defects frequencies are matched to the frequency $\omega^{(2)}_+$ at which the spectral density $J_+(\omega)$ vanishes. (c) $E_N$ versus time for different chain lengths $N=600$ (blue), $N=700$ (red) and $N=800$ (black). The defects frequencies are matched to $\omega^{(2)}_-$ and  $\gamma$ is the same as in (b), while the time is in units of $\omega_\|(N)^{-1}$ (note that to each value of $N$ corresponds a different unit of time). To ensure that the ion spacing in the centre of the chain is the same for each $N$ the trap frequency is modified according to Eq.~\eqref{eq:trapscale}.}
\label{fig:E_N_gamma_zero}
\end{center}
\end{figure*}

This situation can be modelled by two isolated defects whose vibrations are coupled by a spring of strength $\mathcal K\propto 1/R_{12}^3$. If the two defects are initially in a squeezed state, the interaction will lead to beam-splitter dynamics and create two-mode squeezing as a function of time. Unitarity, however, will make this process periodic with a period determined by $\mathcal K$. For a time interval in which the correlations steadily increase, we show the resulting logarithmic negativity for two isolated ions at a distance $d=18$ in Fig.~\ref{fig:E_N_gamma_zero}(a) (dashed line).

If we consider the two ions embedded in a chain of lighter particles, the frequency of the defect ions is separated from the transverse frequencies of the rest of the chain by a gap, and the associated modes are naturally decoupled from the rest of the transverse dynamics and mainly overlap with the localized  normal modes of the chain. There is, however, a remaining finite overlap with non-localized transverse modes of the chain, which leads to a residual coupling. The resulting entanglement dynamics, as shown by the grey line in Figure ~\ref{fig:E_N_gamma_zero}(a) for a chain of $N=800$, demonstrates  that the presence of the transverse environment therefore adds small amplitude oscillations to the unitary entanglement evolution and also tends to decrease its magnitude. 

Let us now assume that at $t=0$ the coupling between the defects transverse oscillations with the axial modes is switched on. This leads to a rapid growth of the quantum correlations between the defect ions and the corresponding logarithmic negativity (see black line in Figure~\ref{fig:E_N_gamma_zero}(a)) and an increase almost tenfold with respect to the value obtained using direct Coulomb coupling only. We note that this kind of dynamics can be observed for any distance larger than $d > 3$. 

It is insightful to compare the results obtained above, which were calculated assuming that the ions form a chain in a linear Paul trap, with the ones found when the ions are assumed to be equally spaced in the axial direction \cite{Fogarty:2013}. The logarithmic negativities for both cases are shown in Fig.~\ref{fig:E_N_gamma_zero}(b) and one can clearly see that the average steady-state entanglement created is comparable, even though slightly reduced in the linear Paul trap setting. However, the revival time $t_\text{rev}$, at which finite-size effects start to play a role, is doubled in the Paul trap, which can be understood by realising that the inter-particle distance at the chain center in the trap is equal to the inter-particle distance in the case of equidistant ions. The increased inter-particle distance far from the chain centre in a harmonic trap suggests that the shorter wavelength modes are localized at the chain center, while the overlap with modes that oscillate away from the centre, is smaller than in the case of a uniform chain. This explains the longer revival time.

Finally, in Fig.~\ref{fig:E_N_gamma_zero}(c) we show the behaviour of the logarithmic negativity for increasing particle number. One can clearly see that the (quasi) steady state value reached does not change as $N$ becomes larger, which allows to extrapolate to the thermodynamic limit and the entanglement reached between two defects embedded in a macroscopic bulk. While the large numbers of ions considered in this plot are so far difficult to achieve in linear-Paul traps, this result shows that entanglement could be observed  at equilibrium in a solid-state environment, provided that spatially-localized normal modes involving the defects characterize the dynamics of the bulk.
  
\subsection{Dependence on the initial state}

\begin{figure}[t]
\begin{center}
\subfigure[]{\includegraphics[width=0.23\textwidth]{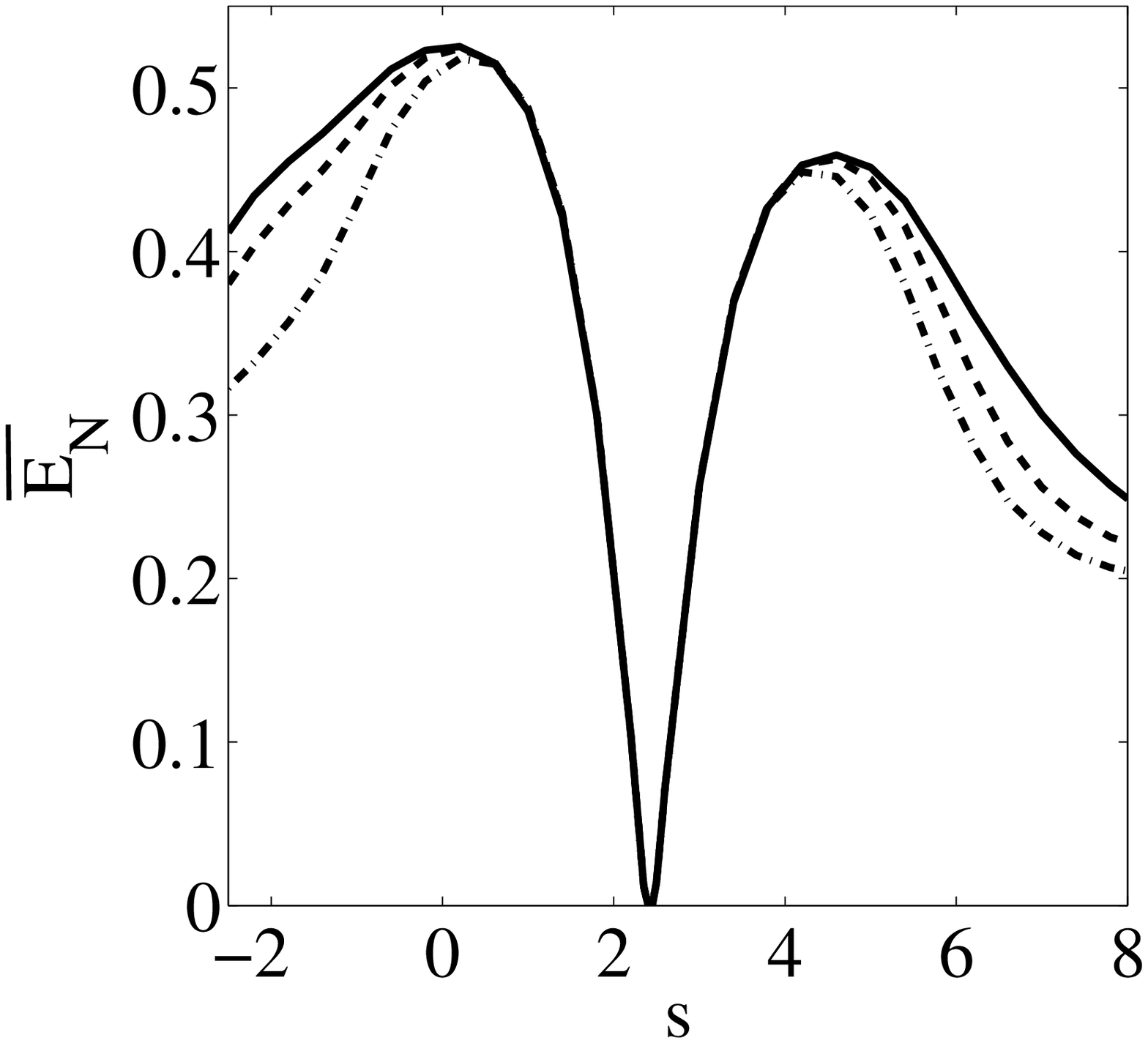}}
\subfigure[]{\includegraphics[width=0.23\textwidth]{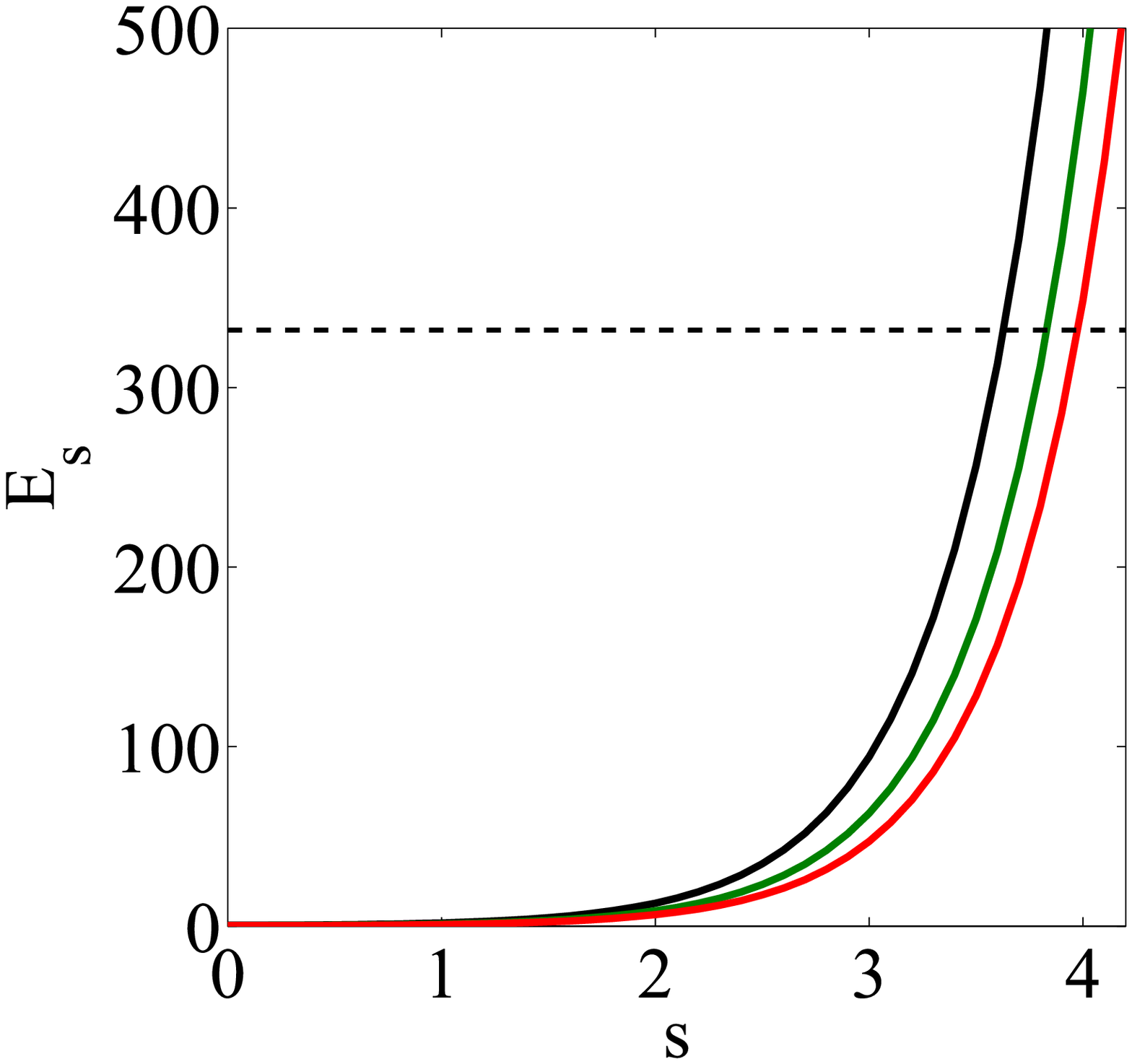}}
\caption{(Color online) (a) Average logarithmic negativity  $\overline{E_N}$ as a function of initial squeezing $s$ of the defects transverse motion. Chain sizes of $N=600$ (dot-dashed line), $900$ (dashed line) and $1200$ (solid line) are shown. (b) The scaled energy imparted by the squeezed states $E_s/N$ as a function of the initial squeezing $s$ for $N=600$ (black), $N=900$ (green) and $N=1200$ (red). The scaled energy of the chains $E_{chain}/N$ is shown for comparison (dashed line). In both figures the spacing is constant between the ions for all $N$ and they are scaled with respect to the minimum spacing between ions in an $N=1200$ ion chain such that $a=2.29 (N)^{-0.596}$. The parameters used are $d=18$ and $\gamma= 4\times 10^3 U_\parallel$ and the frequency of the defects is set to $\omega^{(2)}_-$.}
\label{fig:entmap}
\end{center}
\end{figure}

The results shown so far demonstrate that entanglement can be efficiently generated in the system described above. Two important prerequisites are that the defects are initially prepared in a squeezed state and that the chain is sufficiently cold when the defects are at a distance $d>0$. Let us first examine the dependence on the initial squeezing and assume the chain to be at a sufficiently low temperature.

The dependence of bath-mediated entanglement on the initial squeezing of the entangled objects was already pointed out and discussed in various settings \cite{Paz:2008,Wolf:2011,Kajari:2012,Fogarty:2013}. It can be understood in terms of non-unitary beam splitter dynamics, in which the degrees of freedom of two non-classical input states (which can be separable) mix and thereby generate quantum correlations. When both defects couple to the same position in the chain, a larger value of the initial squeezing leads to larger entanglement. In our model the squeezing of the defects state at $t=0^+$ is the result of the combined action of (i) the initial state preparation (whose squeezing parameter is given by $s$) and (ii) of the quench performed at $t=0$ by switching on the laser (which also gives rise to a sudden quench of the defect potentials). 

In Fig.~\ref{fig:entmap}(a) we display the average logarithmic negativity, $\overline{E_N}$, as a function of the initial squeezing parameter $s$ and for chains of different sizes, assuming the inter-particle distance to be uniform. This ensures that the defects are at the same distance regardless of how many ions are in the chain, allowing for easier comparison of the effect of the chain size on the average entanglement. We first note that the curves all vanish at the same value of $s=:s_0>0$, where the initial squeezing of the defects and the one due to the quench mutually cancel. This effect was also reported in Ref. \cite{Wolf:2011}. At either side of this point $\overline{E_N}$ reaches a maximum and then starts to decrease. This decrease is more pronounced for smaller sizes, and is due to the fact that for the considered parameters the initial energy of the defect oscillators, $E_s=2 (\sinh^2(s)+\frac{1}{2})\hbar\omega_{\perp}$, is comparable to the thermal energy of the rest of the chain, $E_{chain}=\sum_{j=1}^{N}(\langle n(\omega_j)\rangle+\frac{1}{2})\hbar \omega_{j}$, with $\langle n(\omega_j)\rangle=({\rm e}^{\hbar\omega_j/\kappa_BT}-1)^{-1}$. This excess energy causes the second moments of the impurity to undergo large amplitude oscillations which results in large amplitude oscillations of the logarithmic negativity. To avoid this scenario one can place a constraint on the allowed initial squeezing of the defects as 
\begin{equation}
\frac{E_s}{N}\ll\frac{E_{chain}}{N} \, .
\label{eq:sRel}
\end{equation}
In Fig.~\ref{fig:entmap}(b) the scaled energy of the squeezed states (solid lines) is seen to exceed the scaled energy of the chain (dashed line) for $s>3$.  For larger chains the effect of the large squeezing is reduced and one can infer that it becomes negligible in the thermodynamic limit.

The dependence on the initial temperature is intimately related to the distance $d$ between the defects, and we will discuss this dependence in the following.

%%%%%%%%%%%%%%%%%%%%%%%%%%%%%%%%%%%%%%%%%%%%%%%
%%%%%%%%%%%%%%%%%%%%%%%%%%%%%%%%%%%%%%%%%%%%%%%
\subsection{Dependence on the distance}
\label{sec:dist_dependence}
\begin{figure}[t]
\begin{center}
\subfigure[]{\includegraphics[width=0.45\columnwidth]{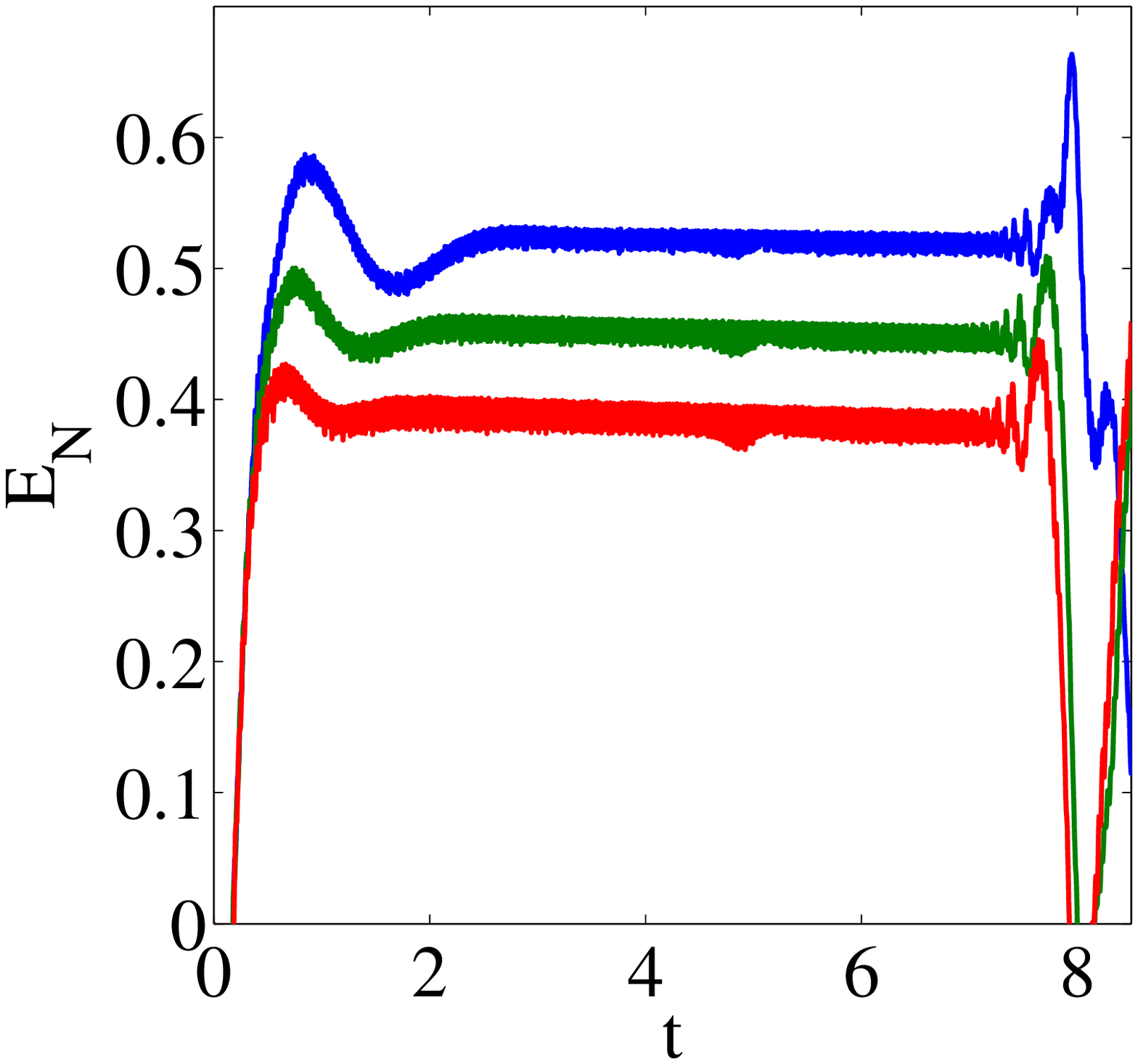}}
\subfigure[]{\includegraphics[width=0.45\columnwidth]{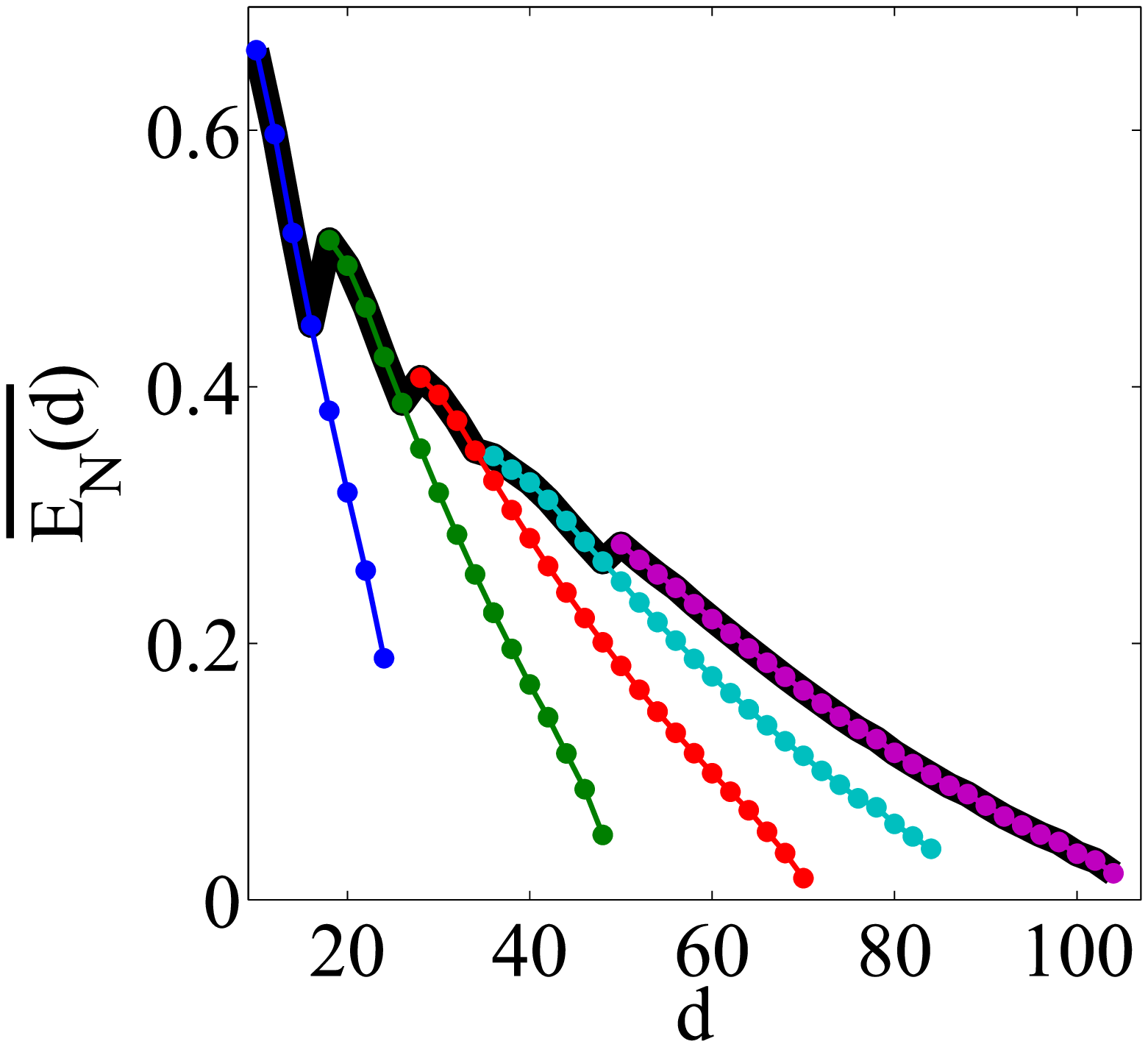}}
%%\subfigure[]{\includegraphics[width=0.66\columnwidth]{m3.eps}}
\caption{(Color online) (a) Logarithmic negativity as a function of time (in units of $\omega_{\parallel}^{-1}$) with varying distances $d=12$ (blue), $14$ (green) and $16$ (red). The parameters are $N = 800$, $T = 0$, $s = 1.5$ and coupling strength $\gamma= 1.9 \times 10^3 U_\parallel$. The defects frequencies are matched to $\omega^{(2)}_-$. (b) Average steady-state entanglement $\overline{E_N}$ as a function of the defects mutual distance. Solid lines connect $\overline{E_N}$ when decoupling via the same $\ell$-th zero of $J^-(\omega)$. $\ell=2$ (blue), $3$ (green), $4$ (red), $5$ (cyan) and $6$ (magenta). The solid black line shows the maximum $\overline{E_N}$.}
\label{fig:dist_dependence}
\end{center}
\end{figure}

To determine the scaling of entanglement as a function of $d$, we numerically determine the logarithmic negativity at the quasi-steady state for a large chain of $N=800$ ions in which the relative motion of the defects contributes to the oscillations of a localized eigenmode (thus, the frequency of the defect oscillators coincide with a value $\omega^{(\ell)}_-$ at which the spectral density $J_-(\omega^{(\ell)}_-)=0$). It can be seen from Fig.~\ref{fig:dist_dependence}(a) that an entangled (quasi) steady state is reached for the cases of $d=12$ (blue), $14$ (green) and $16$ (red curve) and for $\ell=2$, so that $\Omega_{\gamma}=\omega^{(2)}_-$. However, the time at which it is established varies.

The mean value $\overline{E_N}$ of the entanglement over the (quasi-)steady state is shown in  Fig.~\ref{fig:dist_dependence}(b) and can be seen to vary with $d$.  Different colours correspond to different zeros of $J_-(\omega^{(\ell)}_-)$  and show that entanglement production can be optimised by proper choice of the roots of the spectral density, $\omega^{(\ell)}_{\pm}$, to which the defects'  frequency is tuned. The optimal choice of $\ell$ as a function of the distance $d$ between the impurity defects is indicated by the solid black line. Note that for the chosen parameters, no entanglement is present in the steady-state for the first node $\omega^{(1)}_-$, while for any given distance it increases with $\ell$. On the other hand, the time necessary for reaching the steady state also grows with $\ell$, hence for localized normal modes with a large eigenfrequency $\omega^{(\ell)}_-$, the dynamics of entanglement may not exhibit separation of time scales: finite-size effects become evident before a constant value of the entanglement negativity has been reached. This is not necessarily bad. In fact, as observed in Ref. \cite{Plenio:2004} and as we discuss in the next section, finite-size effects can increase the amount of entanglement between the defects, even though for a comparatively shorter interval of time.  

\begin{figure}[h] 
\includegraphics[width=\columnwidth]{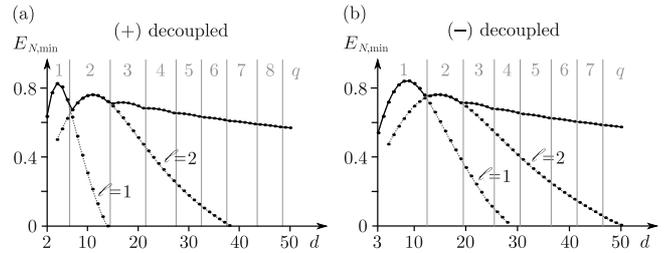}
\caption{Distance dependence of the minimal logarithmic negativity, $E_{N,{\rm min}}=\min(E_N(\bar{t}))$ for $t_\text{th}<\bar{t}<t_\text{rev}$, with the distance $d$ for a chain of coupled oscillators with nearest-neighbour interactions. Subplot (a) shows the behaviour for the  decoupled collective mode of the COM (a) and (b) for the relative motion. The decoupling is achieved with the help of the $\ell$-th zero of the spectral density $J_+(\omega)$ (a) and $J_-(\omega)$ (b), respectively. The parameters chosen are  $m/M=0.5$, $\gamma=0.1\kappa$, $\kappa=1M\Omega_\gamma^2$, see Sec. \ref{micromodel}. The initial temperature is $T=0$ and the squeezing is $s=1$.}
\label{fig:DistanceDependenceTZero}
\end{figure}

Therefore, assuming that one can optimize the dynamics by tuning the frequencies of the defect oscillators such that they match the optimal localized normal mode,  the dependence of the asymptotic value of entanglement on the distance seems to follow a power law behaviour. This can be verified using the model of coupled oscillators with nearest-neigbour interaction, which is discussed extensively in Refs. \cite{Wolf:2011,Kajari:2012} and summarized in Sec. \ref{Sec:MicroModel}. In Fig.~\ref{fig:DistanceDependenceTZero} we show the behaviour of entanglement as a function of the distance when (a) the COM motion and (b) the relative motion of the defects participate in a localized eigenmode. The values of $\ell$ correspond to the roots $\omega^{(\ell)}$ at which the corresponding spectral density vanishes, such that $\Omega_\gamma=\omega^{(\ell)}$. Comparison with the result using the Coulomb interaction demonstrates that the power law decay is found for both the long-range (Coulomb) and the short-range interacting models. It is thus a feature related to the properties of the localized normal modes and their coupling with the rest of the chain. 

%%%%

It is interesting to compare this result with previous studies on bath mediated entanglement in linear chains of oscillators. There, it was shown that entanglement decays quickly on the length scale of the order of the interparticle distance \cite{Audenaert:2002, Anders:2008,Zell:2009}. These works, however, considered chains which possess discrete translational symmetry, while in our case the presence of the defect ions gives rise to a set of localized normal modes involving the defects and the ions between them. These modes are the key elements at the basis of the creation of the decoherence-free subspace which allows for entanglement creation, and show that the dynamics of this entanglement generation is intrinsically related with the lattice symmetries.

%%%%

Further calculations show that the entanglement decay with distance is faster with increasing temperature. In fact, the temperature also determines the variances of the chain oscillators which participate in the localized normal mode, and which are easily identified (in the oscillator chain with nearest-neighbour interactions) as the oscillators interposed between the two defects. The larger the distance between the defects, the larger the number of these oscillators and therefore the more significant  the effect of the chain's temperature, which results from the initial squeezing of the defects and the thermal excitation of the interposed oscillators. Already at $T = 0$, the initial variance of the localized modes is pushed close to the standard quantum limit for large distances. In addition, a number of eigenmodes, which are thermally occupied, have a finite overlap with the defect oscillators and contribute to the variance of both relative and center-of-mass oscillators, thus diminishing the resulting entanglement.

\subsection{Small chains}

\begin{figure}[t]
\begin{center}
\subfigure[]{\includegraphics[width=0.23\textwidth]{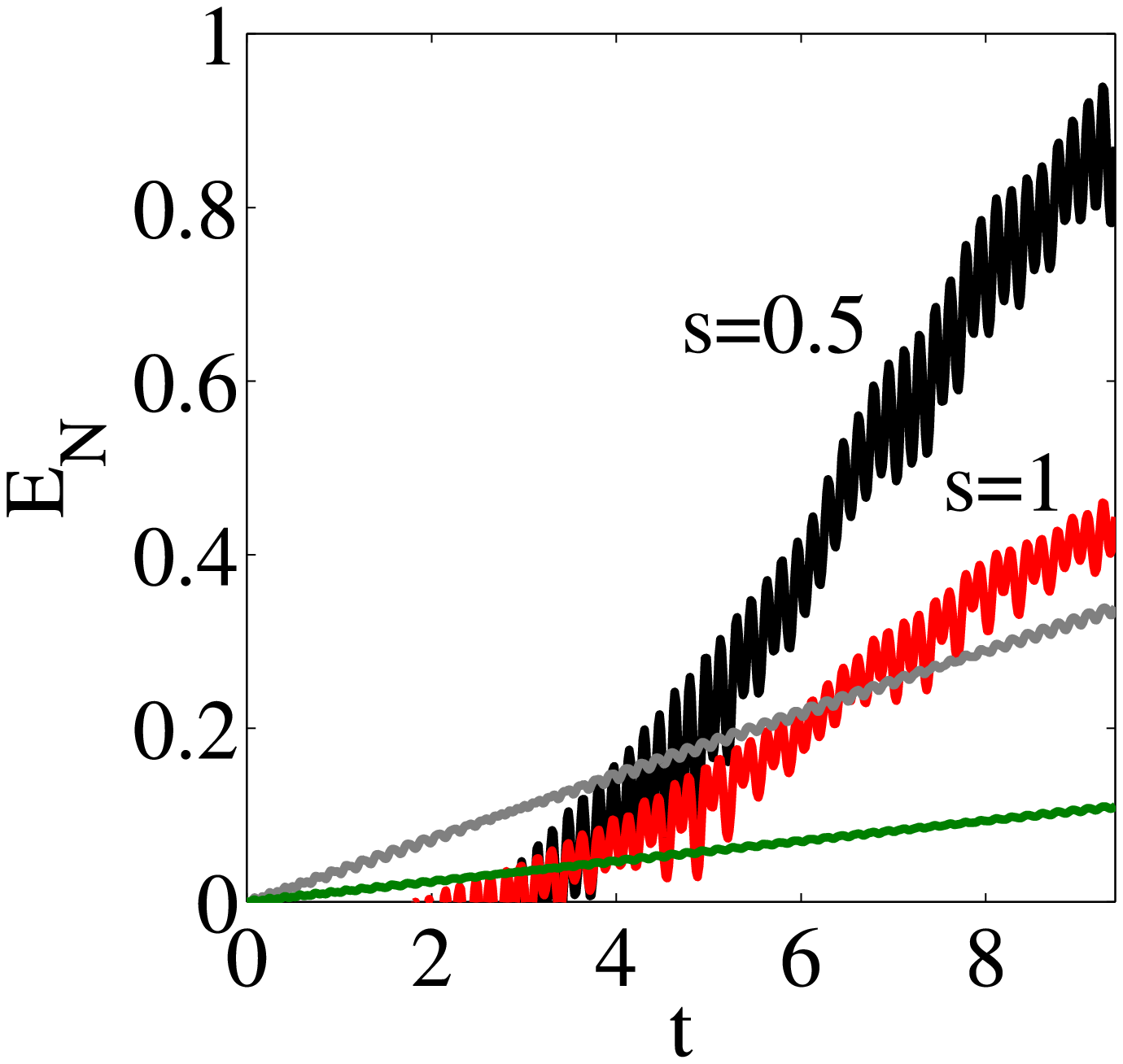}}
\subfigure[]{\includegraphics[width=0.23\textwidth]{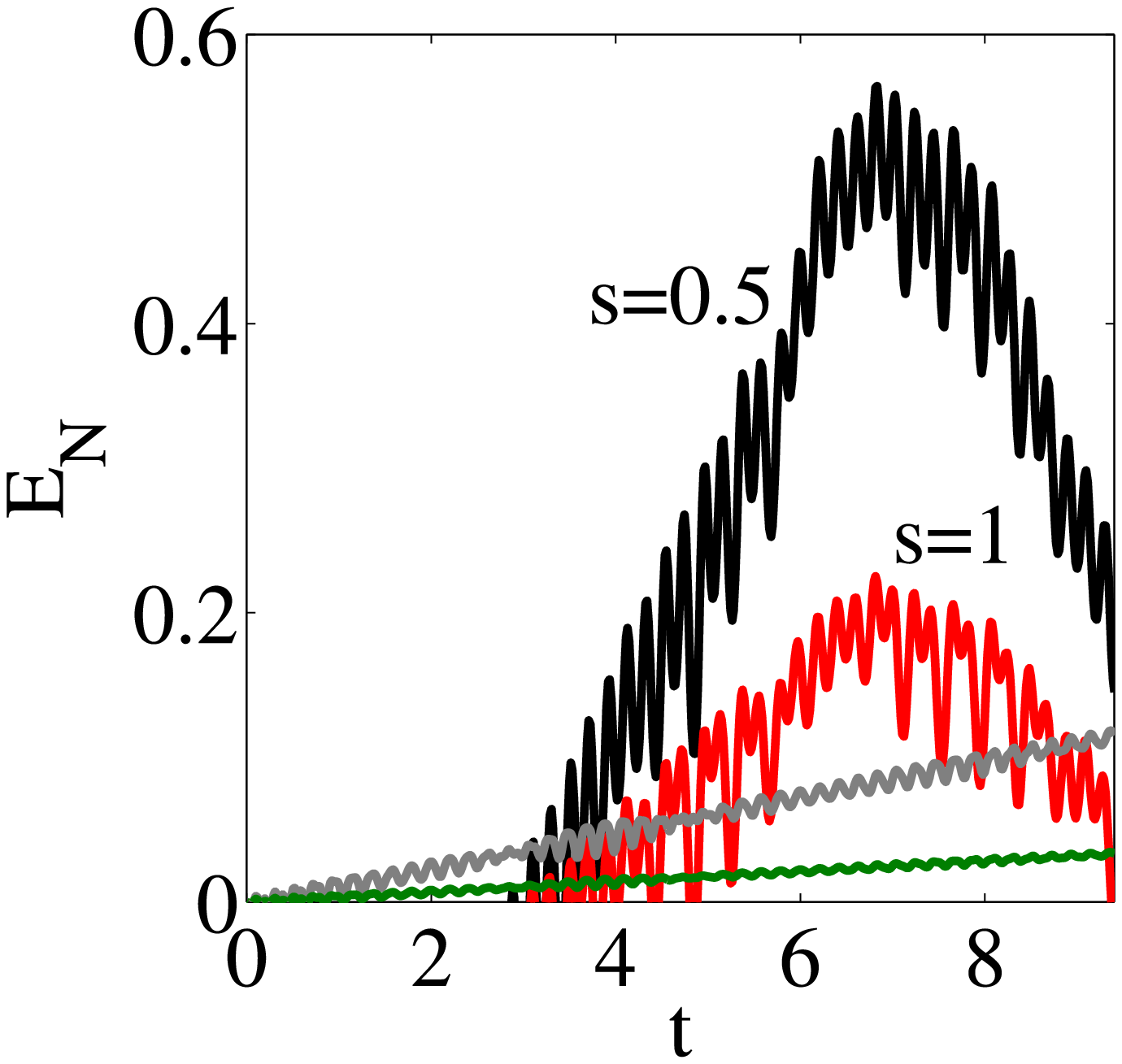}}
\caption{(Color online) Logarithmic negativity as a function of time (in units of $\omega_{\parallel}^{-1}$) for $N = 50$, initial squeezing $s = 0.5$ (black) and $s=1$ (red) while the chain is initially prepared at $T = 8\hbar\omega_\parallel/k_B =10 \mu$K. The coupling strength is $\gamma=13.6 U_\parallel$ and the defects are tuned to $\omega^{(2)}_-$. Panel (a) corresponds to the distance $d=4$ and panel (b) to $d=6$. For comparison the entanglement is also shown when $\gamma=0$ for $s=0.5$ (grey lines) and $s=1$ (green lines).}
\label{fig:E_N_N50}
\end{center}
\end{figure}

We now turn to systems which have been realized in experimental setups consisting of tens of ions forming a chain in a linear Paul trap.  In particular we consider a chain of $N=50$ ions at finite temperatures, in which two defects are embedded at a distance $d=4$ and, in the second case, $d=6$ from each other (i.e. four and six ions, respectively, are interposed). For systems of this size, finite-size effects become relevant on short time scales, which can be seen  from Fig.~\ref{fig:E_N_N50}, where we show the dynamics of the logarithmic negativity for a time scale of the order of the revival time: the average value of entanglement either monotonously increases or decreases and no (quasi) steady-state is reached. In both cases, the coupling with the axial modes gives rise to a tenfold increase of the logarithmic negativity at each instant of time (after a finite transient), when compared to the value due to the direct Coulomb coupling within the transverse chain.

%%%%%%%%%%%%%%%%%%%
%%%%%%%%%%%%%%%%%%%

\begin{figure*}[t]
\begin{center}
\subfigure[]{\includegraphics[width=0.30\textwidth]{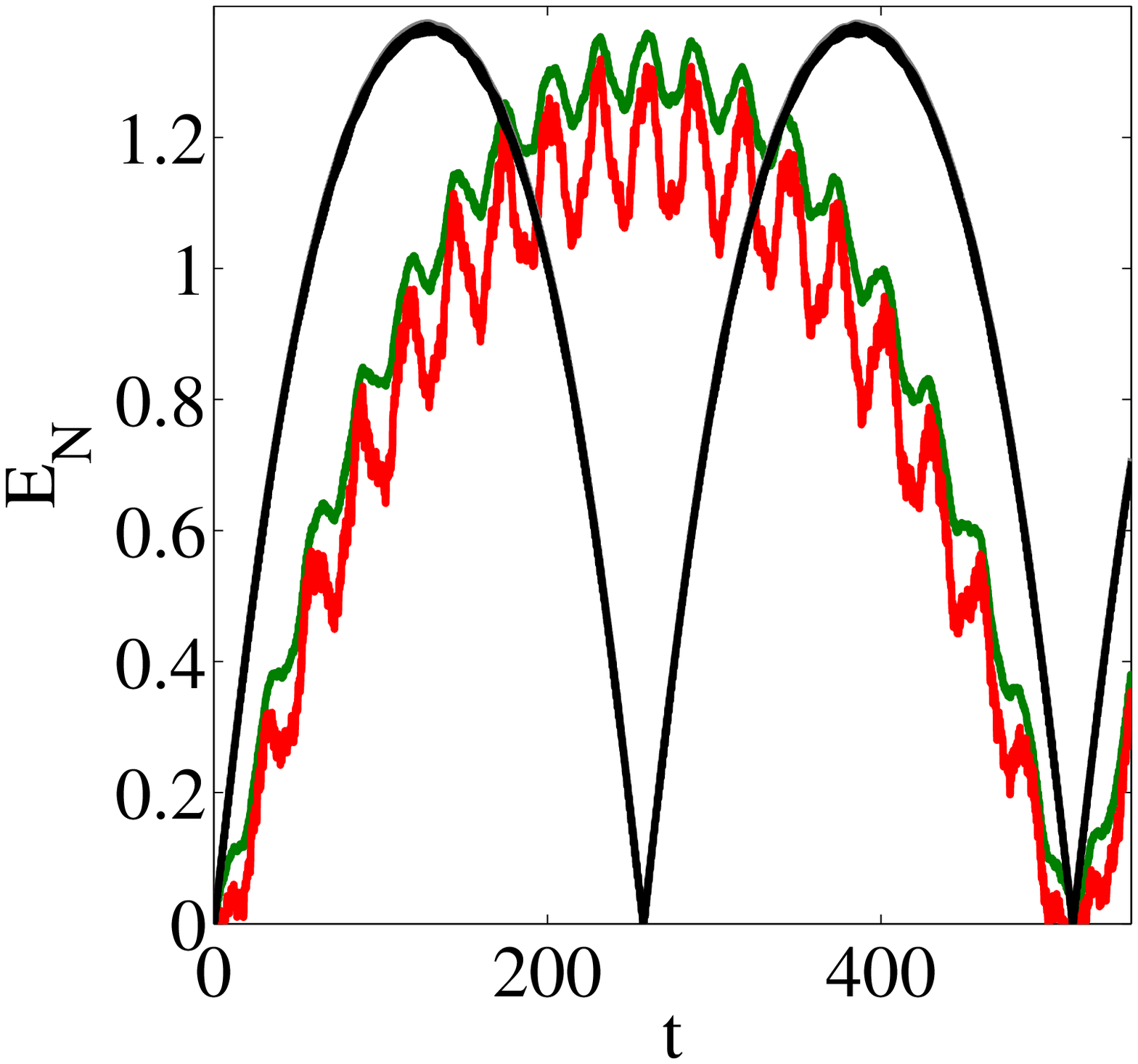}}
\subfigure[]{\includegraphics[width=0.30\textwidth]{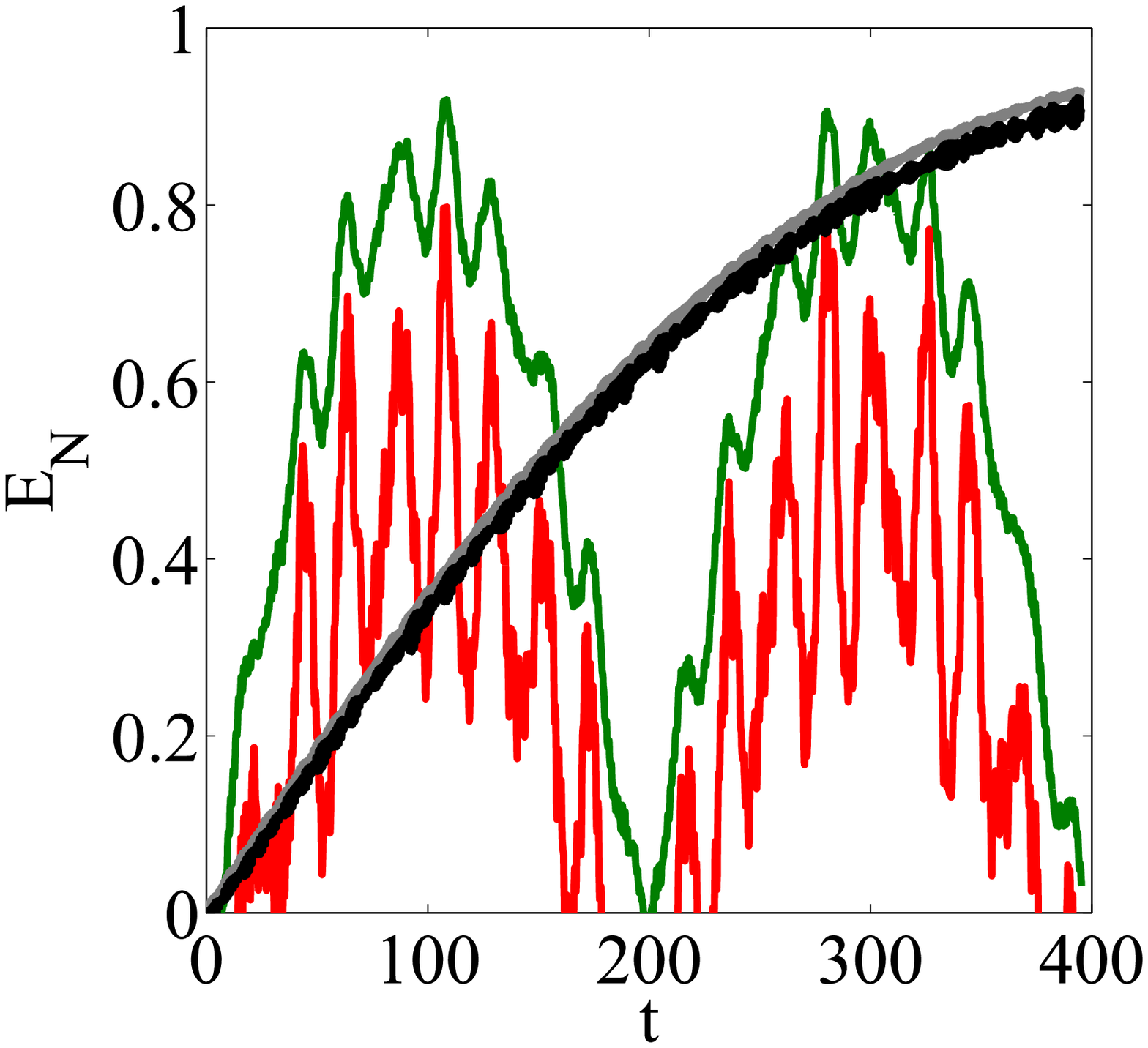}}
\subfigure[]{\includegraphics[width=0.30\textwidth]{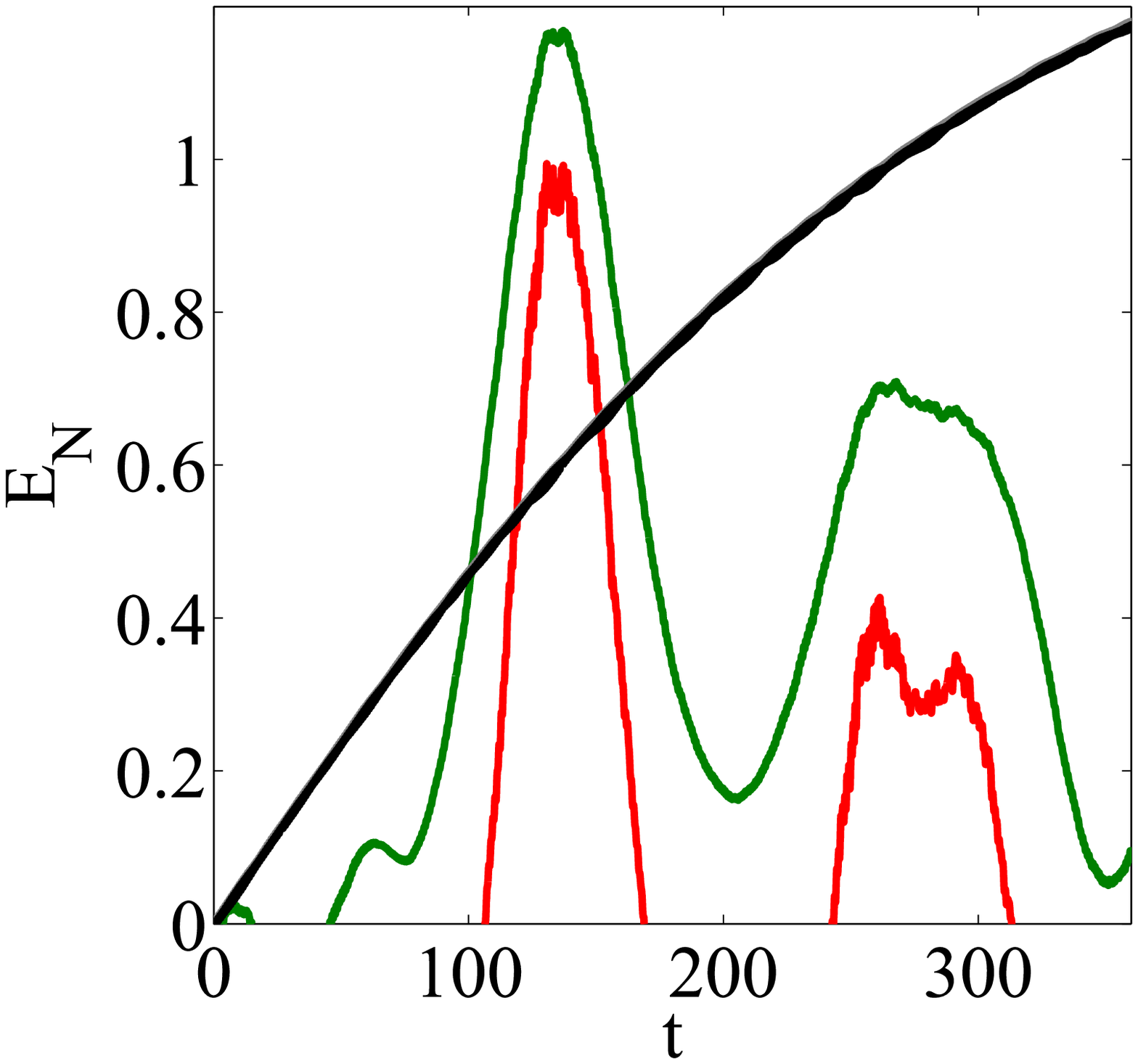}}
\caption{(Color online) Logarithmic negativity as a function of time (in units of $\omega_{\parallel}^{-1}$) for $N = 10$, when the transverse modes have been cooled to the ground state. The chain is initially prepared at temperatures $T = 0.8\hbar\omega_\parallel/k_B =1 \mu$K (green lines) and $T = 4\hbar\omega_\parallel/k_B =5 \mu$K (red lines). The coupling strength is $\gamma=0.5 m\omega_\parallel^2$  in all cases. For comparison the cases for $\gamma=0$ are shown for $T=1\mu$K (grey) and $T=5\mu$K (black). Panel (a) shows $d=2$, panel (b) $d=4$ and the defects are tuned to $\omega^{(1)}_+$. Panel (c) shows $d=4$ for the case where the defects are tuned to $\omega^{(2)}_-$.}
\label{fig:entN10}
\end{center}
\end{figure*}

For chains consisting of $N=10$ ions one can also take advantage of the finite-size effect to generate entanglement. The results are shown in Fig.~\ref{fig:entN10} for $d=2$ and $d=4$ for two different temperatures. For comparison the case of zero laser coupling is shown as grey and black lines for the same temperatures. The oscillatory nature of $E_N$ is clearly visible when the COM motion is decoupled (Figs.~\ref{fig:entN10} (a) and (b)), but is less evident when the relative motion is decoupled (Figure~\ref{fig:entN10}(c)). The coupling with the axial modes changes the frequency of the oscillations of $E_N$ and gives rise to a dynamics similar to the beam-splitter coupling generated by the direct interaction, as can be seen by comparing the curves obtained with and without the coupling laser.

%%%%%%%%%%%%%%%%%%%%%%%%%%%%%%%%%%%%%%%%%%%%%%%
%%%%%%%%%%%%%%%%%%%%%%%%%%%%%%%%%%%%%%%%%%%%%%%
\subsection{Entanglement measurement}
\label{sec:measure}

The quantum states of a single oscillator can be measured by means of an ancillary qubit \cite{Lutterbach1997,Bardroff1999,Zheng2000,Hofheinz2009}, which in the case of an ion chain can be the electronic spin of the ions. The measured statistics of the spin operators then gives direct information of the oscillators quasi-probability functions. Measurements of the bath non-Markovianity have been proposed by means of two qubits \cite{Addis:2013,Borrello:2013}, and specific realizations for ion chains have been discussed. Entanglement measurements, however, present an additional challenge as they are not accessible from single ion statistics and require correlated measurements on the system. Several methods, however, have been developed in recent years \cite{Solano1999,Zheng2000,Hofheinz2009} and here we present a method which extends the single-oscillator state determination introduced in \cite{Tufarelli2011a}. 

To determine the state of the transverse modes of the two defects we assume that one can locally couple spin and motional degrees of freedom of each defect ion by implementing the interaction Hamiltonian $$H^{int}_{i=1,2}=g_i(t)\sigma_i^z(a_i e^{-i\Omega_i t}+a^\dagger_i e^{i\Omega_i t})\,,$$ where $i$ labels the defect ions, $a_i$ ($a^\dagger_i$) is the creation (annihilation) operator for the transverse mode $i$ with frequency $\Omega_i$ and $\sigma_i^{x,y,z}$ the Pauli operators acting on the $i$-th spin. Setting the initial spin state of each ion to be an eigenstate of the $\sigma^x$ operator with eigenvalue $+1$, one can check that, by measuring the expectation value of the correlated spins operators 
$$\langle T \rangle=\langle\sigma^x\otimes\sigma^x-\sigma^y\otimes\sigma^y+i\sigma^x\otimes\sigma^y+i\sigma^y\otimes\sigma^x\rangle\,,$$ one obtains a measurement of the characteristic function $\chi$, which reads
\begin{align}
\chi_\rho(\alpha,\beta)
&=\textmd{Tr}\left[\rho(t)\,D_{j_1}(\alpha)\otimes D_{j_2}(\beta)\right]\\
&=4\langle T \rangle(t)\,.
\end{align}
Here, $\rho(t)$ is the density matrix of the two transverse oscillators at time $t$, $D$ are displacement operators acting on the defects TM, and 
\begin{align}
&\alpha(t)=2i\int_0^{t}dt' g_{j_1}(t')e^{i\Omega_{j_1}t'}\, , \\
&\beta(t)=2i\int_0^{t}dt' g_{j_2}(t')e^{i\Omega_{j_2}t'}\,.
\end{align}
Access to the entire phase space and, consequently, full reconstruction of the oscillators state and its entanglement properties is thus possible by properly designed coupling profiles. For the dynamics of the defect ions, which are Gaussian states and therefore fully described by their low-order moments, it is sufficient to probe their characteristic function close to the phase-space origin. We remark that this scheme demands neither interaction of the distant systems to a common ancilla \cite{Zheng2000}, nor prior displacement operations on the oscillator states \cite{Solano1999,Hofheinz2009}.

%%%%%%%%%%%%%%%%%%%%%%%%%%%%%%%%%%%%%%%%%%%%%%%
%%%%%%%%%%%%%%%%%%%%%%%%%%%%%%%%%%%%%%%%%%%%%%%
\section{Conclusions}
\label{sec:conclusions}

We have theoretically analysed the generation of entanglement between two impurity defects embedded in an ion chain in a linear Paul trap. In this work we have shown that parameter regimes exist for which entanglement can be maintained over large time intervals, with values oscillating around a finite mean-value. The predicted entanglement is robust against fluctuations in the system parameters, such as the size and the elapsed time, as long as finite size effects can be neglected. This therefore suggests that entanglement can be encountered in macroscopic bulk systems as well. While the explored regimes include cases which are  out of reach for current experiments with trapped ions, the entangling dynamics analysed here can be encountered in other physical platforms as well, for example optomechanical systems \cite{Ludwig:2010,Circuit:QED,Marquardt:2012,Houk}. The key ingredients required are the presence of non-local, decoherence-free subspaces (zeroes of the spectral densities) which partially protect the initial information. In addition, the ability to tune the frequencies of the defects to an appropriate zero of the spectral density is required.

Beyond possible applications for quantum technologies, this analysis highlights the conditions for identifying quantum mechanical features in mesoscopic bulks. Important conditions we have identified are the capability to cool the bulk to a very low temperature, which needs to be smaller than the defect frequency, and the ability to tune the defect oscillator frequency to a specific localized normal mode of the bulk. This latter condition requires first a spatial homogeneity of the bulk over the distance between the two defects as well as the possibility to spectrally resolve the individual localized modes eigenfrequencies. These conditions become obviously more and more demanding as the distance is increased, since the frequency distribution of localized eigenmodes  becomes denser and denser within the band. However, this issue could be overcome by applying local time-dependent operations \cite{Kurizki}.

We finally remark that, while the dynamics we have considered here is fully Gaussian, it would be interesting to include non-Gaussian elements, such as  nonlinearities and  initially non-Gaussian states, as these may alter the amount of quantum correlations between the defects \cite{Adesso2007}. Further analysis may also include the effect of the chain close to a structural instability, such as the zigzag transition \cite{Landa:2013}. Here, in fact, modes localised at the chain centre drive the transition and could be used as resources for establishing entanglement between distant points of a network based on solid state devices. 
 
 \acknowledgements The authors are grateful to Cecilia Cormick, Eric Lutz, Alexander Wolf and Gabriele De Chiara for stimulating discussions and useful comments. This work has been financially supported by the European Commission (STREP PICC), by the BMBF (QuORep, Contract No. 16BQ1011), by the German Research Foundation, and by the Okinawa Institute of Science and Technology Graduate University.
%Dave Rea


\begin{thebibliography}{99}

\bibitem{Einstein}
%"ber die von der molekularkinetischen Theorie der Wrme geforderte Bewegung von in ruhenden Flssigkeiten suspendierten Teilchen". 
A. Einstein, Ann. Phys. (Leipzig) {\bf 17}, 549 (1905). 

\bibitem{Chandrasekhar}
%"Stochastic problems in physics and astronomy". 
S. Chandrasekhar,  Rev. Mod. Phys. {\bf 15}, 1 (1943).

\bibitem{FordKacMazur}
%Statistical Mechanics of Assemblies of Coupled Oscillators." 
G. W. Ford, M. Kac, and P. Mazur, J. Math. Phys. {\bf 6}, 504 (1965).

\bibitem{Ullersma}
P. Ullersma, Physica {\bf 32}, 27 (1966); Physica {\bf 32}, 56 (1966); Physica {\bf 32}, 74 (1966); Physica {\bf 32}, 90 (1966).

\bibitem{Rubin:1963}
%Momentum Autocorrelation Functions and Energy Transport in Harmonic Crystals Containing Isotopic Defects
R. J. Rubin, Phys. Rev. {\bf 131}, 964 (1963), and references therein. 

\bibitem{Weiss:1999}
U. Weiss, {\it Quantum Dissipative Systems}, 2nd ed. (World Scientific, Singapore, 1999).

\bibitem{Zurek:2003}
%Decoherence, einselection, and the quantum origins of the classical
W. H. Zurek, Rev. Mod. Phys. {\bf 75}, 715 (2003).

\bibitem{Kinoshita:2004}
T. Kinoshita, T. Wenger, and D. S. Weiss, Nature {\bf 440}, 900 (2006).

\bibitem{Popescu:2006}
%Entanglement and the foundations of statistical mechanics
S. Popescu, A. J. Short, and A. Winter, Nature Phys. {\bf 2}, 754 (2006).

\bibitem{Goldstein:2006}
%Canonical Typicality
S. Goldstein, J. L. Lebowitz, R. Tumulka, and N. Zanghi, Phys. Rev. Lett. {\bf 96}, 050403 (2006).

\bibitem{Rigol:2008}
%Thermalization and its mechanism for generic isolated quantum systems
M. Rigol, V. Dunjko, and M. Olshanii, Nature {\bf 452}, 854 (2008).
 
\bibitem{Riera:2012}
%Thermalization in Nature and on a Quantum Computer
A. Riera, C. Gogolin, and J. Eisert, Phys. Rev. Lett. {\bf 108}, 080402 (2012). 

\bibitem{Schmiedmayer}
M.Gring, M. Kuhnert, T. Langen, T. Kitagawa, B. Rauer, M. Schreitl, I. Mazets, D. Adu Smith, E. Demler, J. Schmiedmayer, Science {\bf 337}, 6100 (2012).

\bibitem{Polkovnikov}
%Colloquium: Nonequilibrium dynamics of closed interacting quantum systems
A. Polkovnikov, K. Sengupta, A. Silva, and M. Vengalattore, Rev. Mod. Phys. {\bf 83}, 863 (2011).
 
\bibitem{Dubin:1999}
D. H. E. Dubin and T. M. O'Neil, Rev. Mod. Phys. {\bf 71}, 87 (1999)

\bibitem{Birkl:1992}
G. Birkl, S. Kassner, H. Walther, Nature {\bf 357}, 310 (1992)

\bibitem{Kajari:2012}
%Statistical mechanics of entanglement mediated by a thermal reservoir
E. Kajari, A. Wolf, E. Lutz, and G. Morigi, Phys. Rev. A {\bf 85}, 042318 (2012).

\bibitem{Wolf:2011}
%Entangling two distant oscillators with a quantum reservoir
A. Wolf, G. De Chiara, E. Kajari, E. Lutz, and G. Morigi, Europhys. Lett. {\bf 95}, 60008 (2011).

\bibitem{Fogarty:2013} T. Fogarty, E. Kajari, B. G. Taketani, A. Wolf, Th. Busch, and G. Morigi, Phys. Rev. A {\bf 87}, 050304 (2013).

\bibitem{EPR}
%Can Quantum-Mechanical Description of Physical Reality Be Considered Complete?
A. Einstein, B. Podolsky, and N. Rosen, Phys. Rev. {\bf 47}, 777 (1935).

\bibitem{Reid}
M.D. Reid, Phys. Rev. A {\bf 40}, 913 (1989); M. D. Reid, P. D. Drummond, W. P. Bowen, E. G. Cavalcanti,
P. K. Lam, H. A. Bachor, U. L. Andersen, and G. Leuchs, Rev. Mod. Phys. {\bf 81}, 1727 (2009).

\bibitem{Duan:2000}
L.-M. Duan, G. Giedke, J. I. Cirac, and P. Zoller, Phys. Rev. Lett. {\bf 84}, 2722 (2000).

\bibitem{Lidar:1998}
%Decoherence-Free Subspaces for Quantum Computation
D. A. Lidar, I. L. Chuang, and K. B. Whaley, Phys. Rev. Lett. {\bf 81}, 2594 (1998). 


\bibitem{MorigiWalther:2002}
G. Morigi and H. Walther, Eur. Phys. J. D {\bf 13}, 261 (2001).

\bibitem{Audenaert:2002}
%Entanglement properties of the harmonic chain
K. Audenaert, J. Eisert, M. B. Plenio, and R. F. Werner, Phys. Rev. A {\bf 66}, 042327 (2002).

\bibitem{Anders:2008}
%Thermal state entanglement in harmonic lattices
J. Anders, Phys. Rev. A {\bf 77}, 062102 (2008). 

\bibitem{Plenio:2004}
M. B. Plenio, J. Hartley, and J. Eisert, New J. Phys. {\bf 6}, 36 (2004).

\bibitem{Huelga:2002}
%Entangled Light from White Noise
M. B. Plenio and S. F. Huelga, Phys. Rev. Lett. {\bf 88}, 197901 (2002);
%An atom that couples to two distinct leaky optical cavities is driven by an external optical white noise field. We describe how entanglement between the light fields sustained by two optical cavities arises in such a situation. The entanglement is maximized for intermediate values of the cavity damping rates and the intensity of the white noise field, vanishing both for small and for large values of these parameters and thus exhibiting a stochastic-resonancelike behavior. This example illustrates the possibility of generating entanglement by exclusively incoherent means and sheds new light on the constructive role noise may play in certain tasks of interest for quantum information processing.

\bibitem{Benatti:2003}
F.~Benatti, R.~Floreanini, and M.~Piani, Phys. Rev. Lett. {\bf 91}, 070402 (2003);
%We show that two, noninteracting two-level systems, immersed in a common bath, can become mutually entangled when evolving according to a Markovian, completely positive reduced dynamics.
%\bibitem{benatti2006}
F.~Benatti and R.~Floreanini, J. Phys. A {\bf 39}, 2689 (2006).
% We consider two independent bosonic oscillators immersed in a common bath, evolving in time with a completely positive, Markovian, quasi-free (Gaussian) reduced dynamics. We show that an initially separated Gaussian state can become entangled as a result of a purely noisy mechanism. In certain cases, the dissipative dynamics allows the persistence of these bath induced quantum correlations even in the asymptotic equilibrium state.


\bibitem{Braun:2002}
D.~Braun, Phys. Rev. Lett. {\bf 89}, 277901 (2002);
% I show that entanglement between two qubits can be generated if the two qubits interact with a common heat bath in thermal equilibrium, but do not interact directly with each other. In most situations the entanglement is created for a very short time after the interaction with the heat bath is switched on, but depending on system, coupling, and heat bath, the entanglement may persist for arbitrarily long times. This mechanism sheds new light on the creation of entanglement. A particular example of two quantum dots in a closed cavity is discussed, where the heat bath is given by the blackbody radiation.


\bibitem{Paz:2008}
J. P. Paz and A. J. Roncaglia, Phys. Rev. Lett. {\bf 100}, 220401 (2008), J. P. Paz and A. J. Roncaglia, Phys. Rev. A {\bf 79}, 032102 (2009).


\bibitem{Zell:2009}
%Distance Dependence of Entanglement Generation via a Bosonic Heat Bath
T. Zell, F. Queisser, and R. Klesse, Phys. Rev. Lett. {\bf 102}, 160501 (2009).

\bibitem{Galve:2010}
%Entanglement dynamics of nonidentical oscillators under decohering environments
F. Galve, G. L. Giorgi, and R. Zambrini, Physical Review A {\bf 81}, 062117 (2010).
 
\bibitem{Mazzola:2009}
L. Mazzola, S. Maniscalco, J. Piilo, K. A. Suominen, and B. M. Garraway, Phys. Rev. A {\bf 80}, 012104 (2009). 

\bibitem{Wolf:2008}
%Assessing Non-Markovian Quantum Dynamics
M. M. Wolf, J. Eisert, T. S. Cubitt, and J. I. Cirac,  Phys. Rev. Lett. {\bf 101}, 150402 (2008).

\bibitem{Breuer:2009}
%Measure for the Degree of Non-Markovian Behavior of Quantum Processes in Open Systems
H. P. Breuer, E. M. Laine, and J. Piilo,  Phys. Rev. Lett. {\bf 103}, 210401 (2009); %Entanglement and Non-Markovianity of Quantum Evolutions
A. Rivas, S. F. Huelga, and M. B. Plenio, Phys. Rev. Lett. {\bf 105}, 050403 (2010).

\bibitem{Vasile:2011}
%Quantifying non-Markovianity of continuous-variable Gaussian dynamical maps
R. Vasile, S. Maniscalco, M. G. A. Paris, H. P. Breuer, and J. Piilo, Phys. Rev. A {\bf 84}, 052118 (2011). 

\bibitem{Liu:2011}
%Experimental control of the transition from Markovian to non-Markovian dynamics of open quantum systems
Bi-Heng Liu, Li Li, Yun-Feng Huang, Chuan-Feng Li, Guang-Can Guo, E.-M. Laine, H. P. Breuer, J. Piilo, Nature phys. {\bf 7}, 931 (2011).

\bibitem{Serafini:2010}
A. Serafini, A. Retzker, and M. B. Plenio, New J. Phys. {\bf 11}, 023007 (2009).

\bibitem{Cormick:2010}
%Observing different phases for the dynamics of entanglement in an ion trap
C. Cormick and J. P. Paz, Phys. Rev. A {\bf 81}, 022306 (2010).


\bibitem{Kielpinski2000a}
%Sympathetic cooling of trapped ions for quantum logic
D. Kielpinski, B. E. King, C. J. Myatt, C. A. Sackett, Q. A. Turchette, W. M. Itano, C. Monroe, D. J. Wineland, and W. H. Zurek, Phys. Rev. A {\bf 61}, 032310 (2000), and references therein.

\bibitem{Dubin:1997}
%Minimum energy state of the one-dimensional Coulomb chain
D. H. E. Dubin, Phys. Rev. E {\bf 55}, 4017 (1997). 

\bibitem{Steane:1998}
A. Steane, Appl. Phys. B: Lasers Opt. {\bf 64}, 623 (1997).

\bibitem{Morigi:2004}
%Eigenmodes and Thermodynamics of a Coulomb Chain in a Harmonic Potential
G. Morigi and S. Fishman, 
Phys. Rev. Lett. {\bf 93}, 170602 (2004); 
%Dynamics of an ion chain in a harmonic potential
Phys. Rev. E {\bf 70}, 066141 (2004).

\bibitem{Drewsen:centralchain}
L. Hornek{\ae}r, N. Kj{\ae}rgaard, A. M. Thommesen, and M. Drewsen, Phys. Rev. Lett. {\bf 86}, 1994 (2001)

\bibitem{Duan}
%Large-scale quantum computation in an anharmonic linear ion trap 
G. D. Lin, S. L. Zhu, R. Islam, K. Kim, M. S. Chang, S. Korenblit, C. Monroe, and L. M. Duan, Europhys. Lett. {\bf 86}, 60004 (2009). 

\bibitem{Champenois:2010}
%An ion ring in a linear multipole trap for optical frequency metrology
C. Champenois, M. Marciante, J. Pedregosa-Gutierrez, M. Houssin, M. Knoop, and M. Kajita, Phys. Rev. A {\bf 81}, 043410 (2010).

\bibitem{Hayasaka2012a}
K. Hayasaka, Appl. Phys. B {\bf 107}, 965 (2012).

\bibitem{Cirac1992}
%Laser cooling of trapped ions in a standing wave
J. I. Cirac, R. Blatt, P. Zoller, and W. D. Phillips,
Phys. Rev. A {\bf 46}, 2668 (1992).

%\bibitem{James1}
%D. F. V. James, Appl. Phys. B {\bf 66}, 181 (1998).


\bibitem{Morigi:2001}
%Doppler cooling of a Coulomb crystal
G. Morigi and J. Eschner, Phys. Rev. A {\bf 64}, 063407 (2001). 

\bibitem{Leibfried:2003}
D. Leibfried, R. Blatt, C. Monroe, and D. Wineland, Rev. Mod. Phys. {\bf 75}, 281 (2003).

\bibitem{Wineland:NIST}
D. J. Wineland, C. Monroe, W. M. Itano, D. Leibfried, B. E. King, and D. M. Meekhof,  J. Res. Natl. Inst. Stand. Technol. {\bf 103}, 259 (1998).

\bibitem{Braunstein2005}
S. L. Braunstein, P. Van Loock, Rev. Mod. Phys {\bf 77}, 513 (2005).



\bibitem{Adesso2007}
G. Adesso and F. Illuminati, J. Phys. A {\bf 40}, 7821 (2007).

\bibitem{Vidal:2002}
G. Vidal and R. F. Werner, Phys. Rev. A {\bf 65}, 032314 (2002); M. B. Plenio, Phys. Rev. Lett. {\bf 95}, 090503 (2005).

\bibitem{Bardroff1999}
P. Bardroff, M. Fontenelle and S. Stenholm, Phys. Rev. A {\bf 59}, R950 (1999)

\bibitem{Hofheinz2009}
M. Hofheinz, H. Wang, M. Ansmann, R. C. Bialczak, E. Lucero, M. Neeley, A. D. O'Connell, D. Sank, J. Wenner, J. M. Martinis and A. N. Cleland, Nature {\bf 459}, 546 (2009)

\bibitem{Zheng2000}
S. B. Zheng, X. W. Zhu, M. Feng, and L. Shi, Phys. Rev. A {\bf 62}, 035801 (2000)

\bibitem{Lutterbach1997}
L. Lutterbach and L. Davidovich, Phys. Rev. Lett. {\bf 78}, 2547 (1997).

\bibitem{Addis:2013}
%Two-qubit non-Markovianity induced by a common environment
C. Addis, P. Haikka, S. McEndoo, C. Macchiavello, and S. Maniscalco, Phys. Rev. A {\bf 87}, 052109 (2013).

\bibitem{Borrello:2013}
%Non-Markovian qubit dynamics induced by Coulomb crystals
M. Borrelli, P. Haikka, G. De Chiara, and S. Maniscalco, Phys. Rev. A {\bf 88}, 010101(R) (2013); 

\bibitem{Solano1999}
E. Solano, R. L. de Matos Filho and N. Zagury, Phys. Rev. A {\bf 59}, R2539 (1999)


\bibitem{Tufarelli2011a}
T. Tufarelli, M. S. Kim and S. Bose, Phys. Rev. A {\bf 83}, 062120 (2011).

%\bibitem{Eberly:2004}
%Finite-Time Disentanglement Via Spontaneous Emission
%T. Yu and J. H. Eberly, Phys. Rev. Lett. {\bf 93}, 140404 (2004).

%\bibitem{Huelga:2012}
%Non-Markovianity-Assisted Steady State Entanglement
%S. F. Huelga, A. Rivas, and M. B. Plenio, Phys. Rev. Lett. {\bf 108}, 160402 (2012). 


\bibitem{Ludwig:2010}
%Entanglement of mechanical oscillators coupled to a nonequilibrium environment.
M. Ludwig, K. Hammerer, and F. Marquardt, Phys. Rev. A {\bf 82}, 012333 (2010).

\bibitem{Circuit:QED}
%Time-reversal symmetry breaking in circuit-QED based photon lattices
%J. Koch, A. A. Houck, K. L. Hur, and S. M. Girvin, Phys. Rev. A {\bf 82}, 043811 (2010); 
%Symmetries and collective excitations in large superconducting circuits
D. G. Ferguson, A. A. Houck, and J. Koch, Phys. Rev. X {\bf 3}, 011003 (2013) 

\bibitem{Marquardt:2012}
%Optomechanical circuits for nanomechanical continuous variable quantum state processing
M. Schmidt, M. Ludwig, and F. Marquardt, New J. Phys. {\bf 14} 125005 (2012).

\bibitem{Houk}
A. A. Houck, H. E. T\"ureci, and J. Koch, Nat. Phys. {\bf 8}, 292 (2012).

\bibitem{Kurizki}
G. Gordon and G. Kurizki, Phys. Rev. Lett. {\bf 97}, 110503 (2006);
J. Clausen, G. Bensky, and G. Kurizki, {\it ibid.} {\bf 104}, 040401 (2010).


\bibitem{Landa:2013}
%Quantum coherence of discrete kink solitons in ion traps
H. Landa, S. Marcovitch, A. Retzker, M. B. Plenio, and B. Reznik, Phys. Rev. Lett. {\bf 104}, 043004 (2010);
%Entanglement Generation Using Discrete Solitons in Coulomb Crystals
H. Landa, A. Retzker, T. Schaetz, B. Reznik, preprint arXiv:1308.2943 (2013).


\end{thebibliography}
\end{document}